\documentclass[journal,twoside]{IEEEtran}
\pdfoutput=1
\IEEEoverridecommandlockouts

\usepackage{color}
\usepackage{times,amsmath,epsfig,amsthm}
\usepackage{amssymb}
\usepackage{subcaption}
\usepackage{comment}
\usepackage{cite}
\usepackage{url}
\usepackage[ruled,linesnumbered]{algorithm2e}
\usepackage{tikz, pgfplots}
\usepackage{moresize}
\usepackage{array}
\usepackage{pgfplots}
\pgfplotsset{compat=1.17}        
\usepackage{caption}             
\usepackage{tikz}                
\usepackage{pgfplotstable}       
\usepackage[linkcolor=black, urlcolor=blue]{hyperref}
\usepackage{booktabs}

\input{mysymbol.sty}
\def \mlp {\mathrm{MLP}}

\usetikzlibrary{shapes,arrows}
\pgfplotsset{compat=1.17}
\pgfplotstableset{col sep=semicolon}
\usepgfplotslibrary{groupplots}
\usepgfplotslibrary{fillbetween}

\tikzset{every mark/.append style={scale=1.6, solid}, font=\small}
\pgfplotsset{
    width=1\textwidth,
    height=5.5cm,
    legend style={
        font=\ssmall ,  
        inner xsep=1pt,
        inner ysep=1pt,
        nodes={inner sep=1pt}},
    legend cell align=left,
    every axis/.append style={line width=.5pt},
 	every axis plot/.append style={line width=1.5pt},
 	every axis y label/.append style={yshift=-4pt}
}

\begin{document}

\title{Directed Acyclic Graph Convolutional Networks}

\author{\IEEEauthorblockN{Samuel Rey, \textit{Member, IEEE,} Hamed Ajorlou, \textit{Student Member, IEEE,} and Gonzalo Mateos, \textit{Senior Member, IEEE}}
\thanks{This work was supported by the NSF under Award ECCS 2231036, by the Spanish AEI Grants PID2022-136887NB-I00 and PID2023-149457OB-I00, and by the Community of Madrid (via grants CAM-URJC F1180 (CP2301), TEC-2024/COM-89, and Madrid ELLIS Unit). Part of the results in this paper appeared at the \textit{2024 Asilomar Conference on Signals, Systems, and Computers}~\cite{dcn2024asilomar}. \emph{(Corresponding author: Gonzalo Mateos)}.

Samuel Rey is with the Dept. of Signal Theory and Communications,
Rey Juan Carlos University, Madrid, Spain (e-mail: samuel.rey.escudero@urjc.es). Hamed Ajorlou and Gonzalo Mateos are with the Dept. of Electrical and Computer Engineering, University of Rochester, Rochester, NY 14627, USA (e-mail: hajorlou@ur.rochester.edu; gmateosb@ece.rochester.edu).}}

\markboth{IEEE Transactions on Signal Processing,~Vol.~XX, June~2025}{Rey \MakeLowercase{et al.}: Directed Acyclic Graph Convolutional Networks}

\maketitle

%
%

\begin{abstract}
Directed acyclic graphs (DAGs) are central to science and engineering applications including causal inference, scheduling, and the automated design of neural architectures. 
In this work, we introduce the DAG Convolutional Network (DCN), a novel graph neural network (GNN) architecture designed specifically for \emph{convolutional} learning from signals supported on DAGs. The DCN leverages causal graph filters to learn nodal representations that account for the partial ordering inherent to DAGs, a strong inductive bias not present in conventional GNNs. Unlike prior art in machine learning over DAGs, DCN builds on formal convolutional operations that admit spectral-domain representations. We further propose the Parallel DCN (PDCN), a model that feeds input DAG signals to a parallel bank of causal graph-shift operators and processes these DAG-aware features using a shared multilayer perceptron. This way, PDCN decouples model complexity from graph size while maintaining satisfactory predictive performance. The architectures' permutation equivariance and expressive power properties are also established. Comprehensive numerical tests across several tasks, datasets, and experimental conditions demonstrate that (P)DCN compares favorably with state-of-the-art baselines in terms of accuracy, robustness, and computational efficiency. These results position (P)DCN as a viable framework for deep learning from DAG-structured data that is designed from first (graph) signal processing principles.
\end{abstract}


%
%

\section{Introduction}

\IEEEPARstart{T}{\lowercase{he}} increasing complexity of data and tasks calls for statistical learning from signals on non-Euclidean domains~\cite{bronstein2017geometric}. Graph Signal Processing (GSP) represents said irregular domains using graphs and leverages their topology as a prior to effectively process relational data~\cite{ortega2018pieee,Dong_2020,leus2023spmag}.
GSP advances have facilitated the principled designs of Graph Neural Network (GNN) architectures and contributed to our understanding of their stability properties, with impact to a host of network-centric machine learning (ML) applications~\cite{bronstein2017geometric,ruiz2021graph,wu2020comprehensive,rozada2025unrolling}.
GNNs learn graph-aware nodal representations through local message passing operations that implement graph convolutions~\cite{kipf2016semi,xu2019powerfulgraphneuralnetworks,ruiz2021graph}, attention mechanisms~\cite{petar2018graphattentionnetworks}, or autoencoders~\cite{wang2017mgae,rey2021overparametrized},
{just to name a few notable instances from an ever-growing list rooted in the seminal recurrent architecture introduced in~\cite{scarselli2008graph}.}


Directed graph models of, e.g., flows, citations, precedence, or, cause-effect relationships are ubiquitous. 
Incorporating directionality into GNNs introduces a strong inductive bias to boost information distillation for scalable learning in this context. However, this comes with well-documented challenges~\cite{marques2020signal}, which are accentuated in the case of \emph{directed acyclic graphs} (DAGs).
{While DAGs do not require iterative message passing due to the absence of cycles, this does not imply that generic aggregation-based GNN architectures are well suited for learning from DAG-supported signals. In fact, designing effective learning models for DAGs remains a nontrivial endeavor, as, e.g., DAGs have nilpotent adjacency matrices, which lead to a collapsed spectrum and prevent the direct use of spectral-based tools~\cite{seifert2023causal,huang2022graph,yang2025deeper}}.
Despite these issues, DAGs are widely adopted in causal inference~\cite{peters2017elements}, learning with Bayesian networks and structural causal models~\cite{peters2014identifiability,zheng2018dags,saboksayr2023colide,dAcunto2023multiscale,rey2025non}, code parsing~\cite{allamanis2018survey}, or performance prediction for architectural optimization of deep neural networks~\cite{zhang2018graph}.

In this work, we introduce a novel \emph{convolutional} GNN specifically designed to learn from signals supported on DAGs; see Section~\ref{ssec:prob_statement} for a formal problem statement. Since DAGs introduce a partial order over the nodes, a central question is how to incorporate this strong inductive bias into the architecture in a principled fashion. Our answer hinges on (and with new results in Section~\ref{ssec:casaulGSO} we contribute to) a recent framework that extends GSP techniques to DAGs and partially ordered sets (posets)\cite{puschel2021discrete,seifert2023causal}. This way, we arrive at a new DAG Convolutional Network (DCN) in Section~\ref{ssec:DCN}. The fundamental design principles adopted (Section~\ref{sec:prelim_prob}) can be traced to linear structural equation models (SEMs), which makes DCN well suited for capturing causal dependencies among variables (provided that the DAG model at hand admits such causal interpretations~\cite{peters2017elements}). In addition, the architecture supports a spectral interpretation and is permutation equivariant. Unlike related GNNs for DAGs\cite{zhang2019dvaevariationalautoencoderdirected,thost2021directed}, DCN's convolutional layers involve mainly sparse matrix multiplications, resulting in manageable computational complexity.

In Section~\ref{ssec:PDCN} we also introduce a Parallel DCN (PDCN) architecture to address potential scalability issues, transitioning from the deep, recursive structure of DCN to a parallel (wide) implementation. Shifted input signals over the DAG are passed through a shared-parameter multilayer perceptron (MLP), and their outputs are summed to form rich nodal representations. We show that the number of learnable parameters in the PDCN model is independent of the graph size, unlike a DCN deployed with the most general class of causal graph filters. We argue that PDCN is permutation equivariant and expressive enough to distinguish between pairs of non-isomorphic DAGs [within the scope of the Weisfeiler-Lehman (WL) test~\cite{WL1968test}].~\vspace{2pt}

\noindent \textbf{Related work and contributions in context.} GNN-based ML over DAGs has been gaining traction recently. A variational autoencoder (VAE) framework called D-VAE was proposed in~\cite{zhang2019dvaevariationalautoencoderdirected}, to embed DAGs into a continuous low-dimensional latent space. 
{D-VAE relies on a sequential encoding of the DAG structure, where node representations are computed following the DAG ordering via recurrent units.}
The DAGNN model in~\cite{thost2021directed} introduces a sequential aggregation scheme, where each node collects information from all its predecessors through an attention mechanism, and then updates its representation using a gated recurrent unit (GRU) to exploit the DAG-induced partial order.
{Consequently, D-VAE and DAGNN 
rely on computationally-intensive sequential operations, 
making them less suitable for large-scale DAGs.}
Instead, DCN captures long-range ancestor dependencies more efficiently via the notion of weighted transitive closure inherent to causal graph filters~\cite{seifert2023causal}. Moreover, DAGNN combines the forward and reverse directions of the DAG in a way that is not straightforward to interpret. More recently,\cite{luo2023transformers} extended transformer models to DAGs by encoding node ordering through a reachability-based attention mechanism. Unlike~\cite{zhang2019dvaevariationalautoencoderdirected, thost2021directed, luo2023transformers}, the proposed (P)DCN architectures build on formal convolutional operators for signals on DAGs. {Accordingly, these new models are best suited for GSP applications where the focus is on ``signal-centric'' predictive tasks, and the DAG is exploited as a strong prior of signal structure.} Our experiments in Section~\ref{sec:experiments} provide evidence in support of this design choice, demonstrating gains in utility for different ML tasks, as well as notable computational savings. Conclusions and a discussion of limitations and future work are given in Section~\ref{sec:conclusion}. 

In summary, this work makes the following significant technical contributions to DAG representation learning:

\begin{itemize}
    \item We develop DCN, a novel architecture that brings to bear causal graph filters for efficient convolutional processing of DAG-structured signals. PDCN trades off depth for width, offering a scalable variant whose parameter count remains constant regardless of the number of DAG nodes.
    \item We examine the representation capabilities of the (P)DCN models, establishing their permutation equivariance and compliance with the WL test. These properties follow from some new results we derive for the causal graph-shift operators and filters pioneered in~\cite{seifert2023causal}.
    
    \item  We assess the performance of the proposed DCN framework on both synthetic and real-world datasets, and show that it compares favorably to state-of-the-art baselines across various tasks. In controlled synthetic experiments, we demonstrate the effectiveness of the model under different noise regimes, number of observations, and graph topologies. We also showcase the practical value of DCN in signal interpolation tasks arising with genomic and hydrological data. We share the code to reproduce all of the results reported in Section \ref{sec:experiments}.
    
\end{itemize}


Relative to the conference precursor~\cite{dcn2024asilomar}, here we consider convolutional learning on DAGs through a unified presentation along with full technical details (including extended discussions, examples, and unpublished theoretical results with their proofs), supported by comprehensive numerical experiments. Noteworthy novel pieces in this journal paper include: (i) the PDCN architecture; (ii) new properties of causal graph-shift operators and filters which are then used to (iii) establish permutation equivariance of (P)DCN as well as examine their expressive power; and (iv) a comprehensive and reproducible performance evaluation protocol.  The latter offers new comparisons with DAGNN, D-VAE, and~DAG+NodeFormer as well as real-data applications that include gene expression level imputation and hydrological data forecasting.

\section{Preliminaries and Problem Statement}\label{sec:prelim_prob}

We start by reviewing the necessary graph-theoretic preliminaries to formally state the problem of learning from DAG signals. We then introduce convolutional GNNs and argue why they are unsuitable as models for the problem dealt with here, which calls for new architectures that are tailored to DAGs. 

%
%

\subsection{Graph-Theoretic Preliminaries: DAGs and Signals}\label{ssec:graph_prelims}

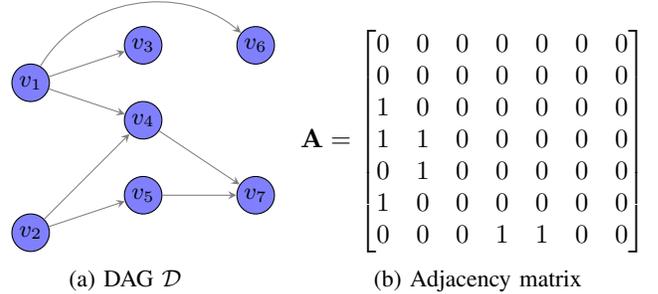
\begin{figure}[t]
\centering

\begin{subfigure}[b]{0.2\textwidth}
\centering
\resizebox{\textwidth}{!}{%
\begin{tikzpicture}[>=stealth]
\hspace{0.3cm}
\tikzset{
    vertex/.style={circle, draw, fill=blue!50, minimum size=5mm, inner sep=0pt},
    edge/.style={draw,->,gray},
}
\node[vertex] (v1) at (0,2) {$v_1$};
\node[vertex] (v2) at (0,0) {$v_2$};
\node[vertex] (v3) at (1.5,2.5) {$v_3$};
\node[vertex] (v4) at (1.5,1.5) {$v_4$};
\node[vertex] (v5) at (1.5,0.5) {$v_5$};
\node[vertex] (v6) at (3,2.5) {$v_6$};
\node[vertex] (v7) at (3,0.5) {$v_7$};

\draw[edge] (v1) -- (v3);  
\draw[edge] (v1) -- (v4);  
\draw[edge, bend left=50] (v1) to (v6);  
\draw[edge] (v2) -- (v4); 
\draw[edge] (v2) -- (v5);  
\draw[edge] (v4) -- (v7);  
\draw[edge] (v5) -- (v7); 
\end{tikzpicture}
}
\caption{DAG $\ccalD$}
\end{subfigure}
\hfill
\begin{subfigure}[b]{0.26\textwidth}
\centering
\resizebox{\textwidth}{!}{%
$
\bbA = 
\begin{bmatrix}
0 & 0 & 0 & 0 & 0 & 0 & 0 \\
0 & 0 & 0 & 0 & 0 & 0 & 0 \\
1 & 0 & 0 & 0 & 0 & 0 & 0 \\
1 & 1 & 0 & 0 & 0 & 0 & 0 \\
0 & 1 & 0 & 0 & 0 & 0 & 0 \\
1 & 0 & 0 & 0 & 0 & 0 & 0 \\
0 & 0 & 0 & 1 & 1 & 0 & 0 \\
\end{bmatrix}
$
}
\caption{Adjacency matrix}
\end{subfigure}

\caption{A DAG $\ccalD$ and its adjacency matrix $\bbA$.}
\label{fig:A_DAG}
\end{figure}

Let $\ccalD = (\ccalV, \ccalE)$ be a DAG, where {$\ccalV = \{v_1,\ldots,v_N\}$} is the set of $N$ nodes and $\ccalE \subseteq \ccalV \times \ccalV$ represents the set of directed edges; see Fig.~\ref{fig:A_DAG} (left).
A set $\ccalV$ with a \emph{partial order} is a set whose elements $v_i, v_j, v_k \in \ccalV$ satisfy: i) $v_i \leq v_i$; ii) $v_i \leq v_j$ and $v_j \leq v_i$ implies $v_i = v_j$; and iii) $v_i \leq v_j$ and $v_j \leq v_k$ implies $v_i \leq v_k$. DAGs and posets are intimately connected: every DAG $\ccalD$ induces a unique partial order on its nodes $\ccalV$, where $v_j < v_i$ whenever there is a directed path from $v_j$ to $v_i$. Conversely, any poset can be realized by at least one DAG whose edges encode the partial ordering relation~\cite{puschel2021discrete}.
In a poset, not all elements are necessarily comparable.
Specifically, for any pair {$v_i, v_j \in \ccalV$} such that there is no path from {$v_i$} to {$v_j$} or from {$v_j$} to {$v_i$}, we have {$v_i \not \leq v_j$} and {$v_j \not \leq v_i$}. Going back to Fig. 1 (left),  e.g., we have {$v_1 < v_7$} while nodes {$v_3$ and $v_4$} cannot be ordered.

Let $\bbA \in \reals^{N \times N}$ denote the (possibly weighted) adjacency matrix of $\ccalD$, where $A_{ij} \neq 0$ if and only if {$(v_i,v_j) \in \ccalE$}.
Given the partial order over $\ccalV$, we can rearrange the entries of $\bbA$ to obtain a strictly lower-triangular matrix; see Fig~\ref{fig:A_DAG} (right).
The diagonal entries are zero, reflecting the fact that DAGs do not contain self-loops.
Beyond the graph structure itself, we also consider signals defined over the node set $\ccalV$.
A graph signal is a function $x: \ccalV \mapsto \reals$, mapping each node to a real value.
This function can be represented as a vector $\bbx \in \reals^N$, where $x_i$ denotes the signal value at node {$v_i$}. For instance, the expression level measured for a gene that is part of gene-regularity network, or, the concentration of a dissolved substance recorded by one of multiple hydrological monitoring stations along a river; see also the numerical experiments in Section \ref{sec:experiments}.
A signal processing framework for graph signals supported on DAGs, referred to as DAG signals, was proposed in~\cite{seifert2023causal}; see also Section~\ref{ssec:DAGSP}.

\subsection{Problem Statement}\label{ssec:prob_statement}
This study focuses on learning from signals defined on a DAG $\mathcal{D}$. The task involves a training dataset $ \ccalT = \{(\bbX_m, \bby_m)\}_{m=1}^M $, comprising $ M $ i.i.d. pairs of input-output observations derived from a network process on $\mathcal{D}$. Each multi-feature input graph signal is represented as $ \mathbf{X}_m \in \reals^{N \times F} $, where $ F $ is the number of input features per node. The outputs $ \bby_m \in \reals^N $ are  single-feature graph signals for simplicity. Extensions to multi-feature output signals or tasks involving graph-level labels are straightforward.

The objective is to learn a mapping between $ \bbX_m $ and $ \bby_m $ using a nonlinear parametric function $f_{\bbTheta}(\cdot ; \ccalD):\reals^{N\times F}\mapsto \reals^N$. Notice how the learnable function is by design chosen to be dependent on the DAG. This way we impose a prior on signal structure to fruitfully exploit the relational inductive biases in the data. The parameters $ \bbTheta $ of this function are optimized by minimizing the empirical risk
\begin{equation}\label{eq:emp_risk_min}
    \min_{\bbTheta} \frac{1}{M} \sum_{m=1}^M \ccalL(\bby_m, f_{\bbTheta}(\bbX_m;\ccalD)),
\end{equation}
where $ \mathcal{L} $ denotes a loss function. For regression tasks, this could be the mean squared error (MSE), while for classification, the cross-entropy loss is commonly used.

%
%

\subsection{Going Beyond Graph Neural Networks}\label{ssec:beyond_gnns}

When it comes to designing the parameteric model $f_{\bbTheta}(\cdot;\ccalD)$, one might be initially tempted to adopt a convolutional GNN off-the-shelf.  
These neural networks employ learnable convolutional graph filters~\cite{elvin2024gf}, where the output of each layer $\ell = 1, \ldots, L$, is recursively given by~\cite{kipf2016semi,ruiz2021graph}:
\begin{equation}\label{eq:fb_gnn}
    \bbX^{(\ell)} = \sigma \left( \sum_{r=0}^{R-1} \bbA^r \bbX^{(\ell-1)} \bbTheta_r^{(\ell)} \right).
\end{equation}
In this architecture, $\bbX^{(\ell)}$ denotes the nodal features at layer $\ell$, while $\bbTheta_r^{(\ell)} \in \reals^{F_i^{(\ell)} \times F_o^{(\ell)}}$ contains the learnable filterbank parameters. $F_i^{(\ell)}$ and $F_o^{(\ell)}$ stand for the number of input and output features. In the initial layer, $\bbX^{(0)}:=\bbX$ is assigned the input graph signals. The powers of the adjacency matrix, $\bbA^r$, determine the radius of the neighborhood from which information is aggregated. This structure enables multi-hop dependencies to be captured by the model. To encode additional graph properties, the adjacency matrix $\bbA$ can be substituted with a more general graph-shift operator (GSO)\cite{ruiz2021graph,rey2025redesigning}, which encodes structural relationships in the graph. Nonlinear transformations are introduced using pointwise activation functions, commonly ReLU, defined as $\sigma(x) = \max(0, x)$. This 
filterbank convolutional GNN (FB-GCNN) has proven effective for scalable learning from network data~\cite{kipf2016semi,ruiz2021graph}.
However, this design is unsatisfactory for the learning problem at hand due to a couple of key reasons. First, the aforementioned graph filters are not convolutional operators for signals on DAGs. They do not satisfy the convolution theorem because the usual GSP notion of Fourier analysis based on GSO eigenvectors is rendered vacuous. Indeed, the strictly lower triangular adjacency matrix $ \bbA $ of a DAG has all eigenvalues equal to zero and, hence, lacks an eigenbasis. While it is operationally feasible to run the GNN model in \eqref{eq:fb_gnn} on a DAG, this collapsing spectrum makes it impossible to have a graph spectral-domain interpretation. 
{Second, DAGs introduce a unique partial order over the set $\ccalV$, which is not explicitly accounted for by generic aggregation mechanisms commonly used in standard GNN architectures.} It is thus prudent to encode this strong inductive bias in the convolutional operation used to learn nodal representations. To address these limitations that motivate our work, we build on the DAG signal processing framework proposed in \cite{seifert2023causal}, which is outlined next.

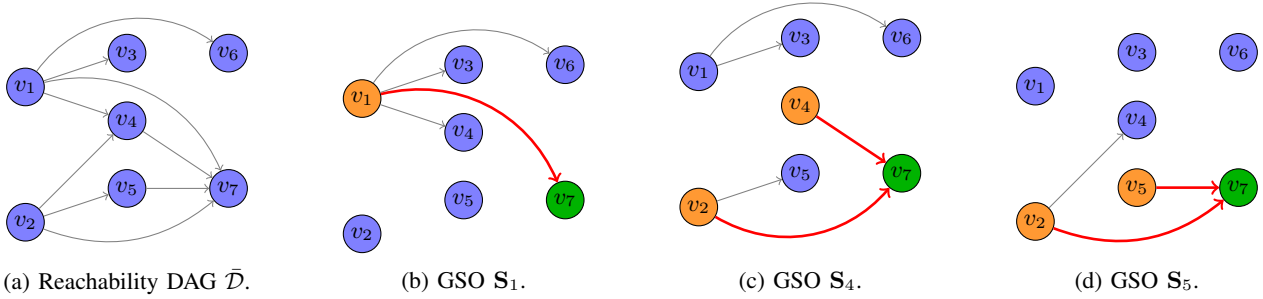
\begin{figure*}[t]
\centering
\tikzset{
    vertex/.style={circle, draw, fill=blue!50, minimum size=5mm, inner sep=0pt},
    selected/.style={vertex, fill=orange!80},
    target/.style={vertex, fill=green!70!black},
    edge/.style={draw,->,gray},
    redge/.style={draw=red,->,line width=1pt}
}

\begin{minipage}[t]{0.24\textwidth}
\centering
\begin{tikzpicture}[->,scale=0.9]
\node[vertex] (v1) at (0,2) {$v_1$};
\node[vertex] (v2) at (0,0) {$v_2$};
\node[vertex] (v3) at (1.5,2.5) {$v_3$};
\node[vertex] (v4) at (1.5,1.5) {$v_4$};
\node[vertex] (v5) at (1.5,0.5) {$v_5$};
\node[vertex] (v6) at (3,2.5) {$v_6$};
\node[vertex] (v7) at (3,0.5) {$v_7$};

\draw[edge] (v1) to (v3);
\draw[edge] (v1) to (v4);
\draw[edge] (v2) to (v5);
\draw[edge] (v2) to (v4);
\draw[edge, bend right=30] (v2) to (v7);
\draw[edge, bend left=45] (v1) to (v6);
\draw[edge, bend left=40] (v1) to (v7);
\draw[edge] (v4) to (v7);
\draw[edge] (v5) to (v7);
\end{tikzpicture}
\subcaption{Reachability DAG $\bar{\ccalD}$.}\label{fig:example_W}
\end{minipage}
%
\begin{minipage}[t]{0.24\textwidth}
\centering
\begin{tikzpicture}[->,scale=0.9]
\node[selected] (v1) at (0,2) {$v_1$};
\node[vertex] (v2) at (0,0) {$v_2$};
\node[vertex] (v3) at (1.5,2.5) {$v_3$};
\node[vertex] (v4) at (1.5,1.5) {$v_4$};
\node[vertex] (v5) at (1.5,0.5) {$v_5$};
\node[vertex] (v6) at (3,2.5) {$v_6$};
\node[target] (v7) at (3,0.5) {$v_7$};

\draw[edge, bend left=45] (v1) to (v6);
\draw[edge] (v1) -- (v4);
\draw[edge] (v1) -- (v3);
\draw[redge, bend left=40] (v1) to (v7);
\end{tikzpicture}
\subcaption{GSO $\bbS_1$.}
\label{fig:example_T1}
\end{minipage}
%
\begin{minipage}[t]{0.24\textwidth}
\centering
\begin{tikzpicture}[->,scale=0.9]
\node[vertex] (v1) at (0,2) {$v_1$};
\node[selected] (v2) at (0,0) {$v_2$};
\node[vertex] (v3) at (1.5,2.5) {$v_3$};
\node[selected] (v4) at (1.5,1.5) {$v_4$};
\node[vertex] (v5) at (1.5,0.5) {$v_5$};
\node[vertex] (v6) at (3,2.5) {$v_6$};
\node[target] (v7) at (3,0.5) {$v_7$};

\draw[edge, bend left=45] (v1) to (v6);
\draw[edge] (v1) to (v3);
\draw[edge] (v2) to (v5);
\draw[redge] (v4) to (v7);
\draw[redge, bend right=40] (v2) to (v7);
\end{tikzpicture}
\subcaption{GSO $\bbS_4$.}
\label{fig:example_T4}
\end{minipage}
%
\begin{minipage}[t]{0.24\textwidth}
\centering
\begin{tikzpicture}[->,scale=0.9]
\node[vertex] (v1) at (0,2) {$v_1$};
\node[selected] (v2) at (0,0) {$v_2$};
\node[vertex] (v3) at (1.5,2.5) {$v_3$};
\node[vertex] (v4) at (1.5,1.5) {$v_4$};
\node[selected] (v5) at (1.5,0.5) {$v_5$};
\node[vertex] (v6) at (3,2.5) {$v_6$};
\node[target] (v7) at (3,0.5) {$v_7$};

\draw[edge] (v2) to (v4);
\draw[redge] (v5) to (v7);
\draw[redge, bend right=30] (v2) to (v7);
\end{tikzpicture}
\subcaption{GSO $\bbS_5$.}
\label{fig:example_T5}
\end{minipage}

\caption{An example of how information is propagated by causal shifts $\bbS_k$ towards downstream nodes. For the GSOs $\bbS_1$, $\bbS_4$, and $\bbS_5$, we indicate the nodes whose signal values contribute to the respective shifted signals at node $v_7$. Self-loops in the directed graphs induced by the GSOs are excluded to avoid clutter.}
\label{fig:example}
\end{figure*}~\label{fig:Example}

%
%

\section{DAG Signal Processing} \label{ssec:DAGSP}
Here we review convolutions for signals on DAGs~\cite{seifert2023causal}, which are fundamental building blocks in our new architectures. We also establish some new properties of DAG convolutional filters that we will use for subsequent analyses.

\subsection{Graph-Shift Operators and Convolution for DAGs}



Consider a signal $\bbx \in \reals^N$ defined in the partially-ordered nodes of a DAG $\ccalD$. The seminal work in~\cite{seifert2023causal} advocates modeling $x_i$ as linearly aggregated (possibly causal) effects $\{x_j:{(v_i,v_j)}\in\ccalE\}$ from parent nodes, and hence all of node {$v_i$}'s predecessors {$v_k < v_i$} by transitivity. Let $\bbc \in \reals^N$ denote the vector of exogenous nodal contributions, where input $c_i$ is associated with node {$v_i\in\ccalV$} following our graph signal terminology. Accordingly, the relationship between the signal $\bbx $ and the contributions $\bbc$ is given by
\begin{equation}\label{eq:signal_model}
x_i=\sum_{j\leq i}W_{ij}c_j, \quad {v_i}\in\ccalV\quad \Rightarrow \quad \mathbf{x} = \mathbf{W}\mathbf{c},
\end{equation}
where $ \bbW \in \reals^{N \times N} $ is the so-termed \emph{weighted transitive closure} matrix of the DAG. An entry $W_{ij} \neq 0 $ indicates that there exists a directed path from node {$v_j$} to node {$v_i$}. 
{Note that, with a slight abuse of notation, in \eqref{eq:signal_model} we write $j \leq i$ to denote the ordering between nodes $v_j$ and $v_i$.}
These {directed paths} are also the edges in the reachability graph $\bar{\ccalD}$ associated with $\ccalD$; see also Fig.~\ref{fig:example_W} for the reachability graph of the DAG in Fig.~\ref{fig:A_DAG}. It follows that matrix $\bbW$ is lower-triangular, and (by design) its diagonal entries are $W_{ii}:=1$  to account for the reflexivity property of posets. In terms of the signal model \eqref{eq:signal_model}, this means that $c_i$ contributes to $x_i$. Still, the most important implication of $\textrm{diag(\bbW)}=\mathbf{1}$ is that $\bbW$ is invertible, and its columns will play the role of a Fourier basis as discussed later. 

We still have not addressed how to compute the non-diagonal entries $W_{ij}$ from those in the adjacency matrix $\bbA$. There are multiple ways to do so depending on the meaning of the edge weights in $\ccalD$; see~\cite[Section III-B]{seifert2023causal} for a detailed treatment and algorithm building on~\cite{lehmann1977algebraic}. Here, we adopt
\begin{equation}\label{eq:trans_closure}
\mathbf{W} = \mathbf{I}+\bbA+\bbA^2+\ldots+\bbA^{N-1}= (\mathbf{I} - \mathbf{A})^{-1},    
\end{equation}
which implies a linear SEM for $\bbx$, when $\bbc$ are exogenous noises~\cite[Ch. 6.2]{peters2017elements}. Indeed, using \eqref{eq:signal_model} and \eqref{eq:trans_closure} yields
$\bbx=\bbW\bbc = (\mathbf{I} - \mathbf{A})^{-1}\bbc=\bbA\bbx+\bbc.$
{Because for DAGs the adjacency matrix $\bbA$ is nilpotent, the inverse $(\bbI-\bbA)^{-1}$ always exists and is numerically stable.}

For each node {$ v_k \in \ccalV $}, a causal GSO $\bbS_k \in \reals^{N \times N} $ can be defined as~\cite{seifert2023causal}
\begin{equation}\label{eq:causal_GSO}
[\bbS_k \bbx]_i = \sum_{j \leq i \text{ and } j \leq k} W_{ij} c_j.
\end{equation}
This operation essentially filters the signal at node {$ v_i $} by aggregating contributions from nodes that are \emph{common} predecessors of {$ v_i $} and {$ v_k $} [cf. \eqref{eq:signal_model}]. In matrix form, \eqref{eq:causal_GSO} can be expressed as
\begin{equation}\label{eq:compact_gso}
    \bbS_k \bbx = \bbW \bbD_k \bbc = \bbW \bbD_k \bbW^{-1} \bbx,
\end{equation}
where $ \bbD_k $ is a diagonal matrix such that $ [\bbD_k]_{ii} = 1 $ if {$ v_i \leq v_k $} and $ [\bbD_k]_{ii} = 0 $ otherwise.
{The diagonal support of $\bbD_k$ corresponds to the set of ancestors of $v_k$, which is typically small for nodes close to the roots. In the extreme case of a root node $v_k$, $\bbD_k$ contains a single nonzero entry $ [\bbD_k]_{kk} = 1 $. This structural sparsity propagates to $\bbS_k$.}
As argued earlier, $ \bbW^{-1} $ is guaranteed to exist and can be efficiently computed using the weighted Möbius inversion~\cite{rota1964foundations}. Inspecting the left- and right-hand side of \eqref{eq:compact_gso}, it follows that the columns of $ \bbW $ are the eigenvectors of all GSOs $ \bbS_k $, and the binary diagonal entries of $ \bbD_k $ the respective eigenvalues. Consequently, $ \bbW $ provides a Fourier basis for DAG signals, while $ \bbW^{-1} $ represents the corresponding Fourier transform. From \eqref{eq:signal_model}, the contributions $ \bbc $ can be interpreted as the Fourier coefficients of $\bbx$.

In GSP, shift-invariant graph filters are polynomials of a single GSO~\cite{elvin2024gf}. However, for DAGs there is one causal  GSO $ \bbS_k $ for every {$ v_k \in \ccalV $}, and each $ \bbS_k $ is idempotent. Accordingly, the most general form of a linear shift-invariant convolutional filter for DAGs (henceforth called causal graph filters) was introduced in~\cite{seifert2023causal} as
\begin{equation}\label{eq:causal_gf}
    \bbH = \sum_{k \in \ccalV} \theta_k \bbS_k = \bbW  \sum_{k \in \ccalV} \theta_k \bbD_k  \bbW^{-1},
\end{equation}
where $\bbtheta=[\theta_1,\ldots,\theta_N]^\top\in\reals^{N}$ are the filter coefficients.
{With a slight abuse of notation, in \eqref{eq:causal_gf} we identify $k \in \ccalV$ with $v_k \in \ccalV$.}
For an input signal $\bbx$, we say $\bbH$ implements the 
convolution operation $\bby=\bbh *_{\ccalD} \bbx = \bbH\bbx$. 
From \eqref{eq:causal_gf} frequency response of the causal graph filter is given by the diagonal of $ \sum_{k \in \ccalV} \theta_k \bbD_k $, thus implementing a pointwise scaling operation on the frequency coefficients $\bbc$ of the input signal $\bbx$. 

This causal graph filter definition forms the foundation of the DCN architecture we introduce in Section \ref{ssec:DCN}. But before moving on, next we provide further insights and establish new properties of causal GSOs and filters that will inform subsequent analyses of the proposed architectures.

%
%

\subsection{Properties of Causal GSOs and Graph Filters}~\label{ssec:casaulGSO}

Let us begin with the illustrative example in Fig.~\ref{fig:example} to develop intuition about how the matrices $\bbS_k$ diffuse information across $\ccalD$. Recall the toy DAG in Fig.~\ref{fig:A_DAG}. Its reachability DAG $\bar{\ccalD}$ is shown in Fig.~\ref{fig:example_W}, which corresponds to the support of the non-diagonal entries in $\bbW$. Indeed, $\bbW$ also includes self-loops at every node. The other panels depict the connectivity of the directed graphs induced by the GSOs $\bbS_1$, $\bbS_4$, and $\bbS_5$. These graphs are not DAGs, because the $\bbS_k$ have some non-zero diagonal elements corresponding to direct and indirect ancestors of $k$ (including itself; these self-loops are not shown in Fig.~\ref{fig:example} to avoid clutter).
For example, if we are interested in the value at node $v_7$ after shifting a signal $\bbx$ with respect to node $v_4$ using the GSO $\bbS_4$, we obtain from Fig.~\ref{fig:example_T4} that
\begin{equation}
   y_7 = [\bbS_4]_{72}x_2 + [\bbS_4]_{74}x_4 + x_7.
\end{equation}
A closer inspection of the connectivity in Fig.~\ref{fig:example} suggests that using the causal GSO to aggregate information at a particular node ($v_7$ in the previous example) only involves a subset of its ancestors and possibly itself. This intuition is formalized in the following result, which establishes a relationship between the supports of $\bbW$ and $\bbS_k$. 

\begin{proposition}\label{prop:same_support}
Consider the weighted transitive closure $\bbW \! = \! (\bbI \!-\! \bbA)^{-1}$.
    For every {$v_k \in \ccalV$}, it holds that $\sup (\bbS_k) \subseteq \sup (\bbW)$.
\end{proposition}
\begin{proof}
The off-diagonal support of $\bbW$ corresponds to the edge set of the reachability DAG $\bar{\ccalD}$, {meaning that $W_{ij} \neq 0$ only if there exists a directed path from {$v_j$} to {$v_i$} in $\ccalD$.}  
By definition, the causal GSOs are given by  
\begin{equation}
        \bbS_k = \bbW \bbD_k \bbW^{-1} = \bbW \bbD_k (\bbI - \bbA).
    \end{equation}
{Since $\bbD_k$ is a diagonal matrix with binary $\{0,1\}$ entries, it only nulls selected columns and therefore cannot create new nonzero entries.
Consequently, if $[\bbW \bbD_k (\bbI-\bbA)]_{ij} \neq 0$, then necessarily $[\bbW (\bbI-\bbA)]_{ij} \neq 0$.
Hence, it suffices to establish the inclusion for the case $\bbD_k = \bbI$, yielding}
    \begin{equation}
        [\bbS_k]_{ij} = \sum_{v \in \ccalV} W_{iv} [\bbI - \bbA]_{vj}.
    \end{equation}
If $[\bbS_k]_{ij} \neq 0$, then there exists $v$ for which both $W_{iv} \neq 0$ and $[\bbI - \bbA]_{vj} \neq 0$.  
This implies that there is a direct link from {$v_j$} to {$v_v$} and a directed path from {$v_v$} to {$v_i$}, meaning that there must also be a path from {$v_j$} to {$v_i$}.  
Hence, $W_{ij} \neq 0$.  Conversely, if $W_{ij} = 0$, then no such {$v_v$} exists, ensuring that $[\bbS_k]_{ij} = 0$. 
\end{proof}

Our goal is to leverage $\bbS_k$ to aggregate information via convolutional operations on DAGs, as illustrated in the previous example.
With $\ccalS \! = \! \{ \bbS_k\}_{k \in \ccalV}$ {denoting the} set of causal GSOs, a natural question that arises is whether different DAGs can generate the same set $\ccalS$.
The answer is negative, as stated in the following proposition.
\begin{proposition}\label{prop:uniqueness}
The set of causal GSOs $\ccalS = \{ \bbS_k \}_{k \in \ccalV}$ derived from the weighted transitive closure $\bbW = (\bbI - \bbA)^{-1}$ are unique, i.e., distinct (non-isomorphic) DAGs yield different matrices $\bbS_k$.
\end{proposition}

The proof is provided in Appendix~\ref{a:uniqueness}.
Proposition~\ref{prop:uniqueness} ensures that the GSOs $\bbS_k$ define a unique aggregation mechanism within a DAG, which is crucial for analyzing the expressivity of the proposed architectures.
Additionally, a key property in learning from graph-based data is permutation equivariance, which ensures robustness to node relabeling.
Graph filters and GNNs possess this property~\cite{ruiz2021graph,rey2025redesigning}, and we now show that causal graph filters do as well; see Appendix \ref{a:equivariance} for a proof.
\begin{theorem}\label{thm:gf_equivariant}
Let $\ccalV'$ be the poset of nodes obtained by permuting with $\pi:\ccalV\mapsto \ccalV'$ the DAG with node ordering $\ccalV$. Let $\bbP$ be the matrix representation of $\pi$. For any {$v_q \in \ccalV'$}, the permuted GSO $\bbS_q'$ is given by
\begin{equation} \label{eq:gso_perm}  
\bbS_q' = \bbP \bbS_k \bbP^\top,  
\end{equation}  
where {$v_k=\pi^{-1}(v_q)$}. Moreover, causal graph filters are permutation equivariant, meaning filter outputs given by $g(\bbx; \bbtheta, \ccalS) = \sum_{k \in \ccalV} \theta_k \bbS_k \bbx$ satisfy
\begin{equation}
       g(\bbx'; \bbtheta', \ccalS') = \bbP g(\bbx; \bbtheta, \ccalS),
\end{equation}
where $\ccalS \! = \! \{ \bbS_k\}_{k \in \ccalV}$ is the set collecting all the GSOs, $\bbx' \!=\! \bbP\bbx$, $\bbtheta' \!=\! \bbP \bbtheta$, and $\ccalS' \!=\! \{ \bbS'_q \}_{q \in \ccalV'}$.
\end{theorem}

\begin{figure*}[t]
    \centering
    \begin{minipage}{0.48\textwidth}
        \centering
        \begin{tikzpicture}[
            scale=0.9,
            ->, >=stealth',
            node distance=1.1cm and 1.4cm,
            every node/.style={font=\small},
            main/.style={rectangle, draw, fill=gray!10, minimum size=0.2cm, text centered},
            source/.style={rectangle, draw, fill=gray!30, minimum size=0.8cm, text centered}
        ]
        
        \coordinate (A) at (-2,-1);
        \coordinate (B) at (3,-1);
        \draw[fill=black] (A) circle (1pt); 
        \draw[fill=black] (B) circle (1pt); 
        \node (x0) at (-4.2,0.5) {\small\(\bbX^{(0)}\)};
        \node (xl) at (-0.5,1) {\small\(\mathbf{X^{(\ell)}}\)};
        \node (xl1) at (2,1) {\small\(\mathbf{X^{(\ell+1)}}\)};
        \node (xl2) at (-2.2,-1.5) {\small\(\mathbf{X^{(\ell)}}\)};
        \node (xl3) at (3.8,-1.5) {\small\(\mathbf{X^{(\ell+1)}}\)};

        \node[source] (DCN1) [minimum size=0.5cm, rounded corners=6pt] at (-2.5,0.5) {DCN layer};
        \node at (-1,0.5) (dots1) {\small$\cdots$};
        \node[source] (DCNl1) [minimum size=0.5cm, rounded corners=6pt] at (0.5,0.5) {DCN layer};
        \node at (2,0.5) (dots2) {\small$\cdots$};
        \node[source] (DCNL) [minimum size=0.5cm, rounded corners=6pt] at (3.5,0.5) {DCN layer};
        \node[main] (DCNl) [minimum size=0.6cm, rounded corners=6pt] at (0,-1.5) {$\sum_{k \in \mathcal{V}} \bbS_k \bbX^{(\ell)}\bbTheta^{(\ell)}_k$};
        \node[main] (sig) [minimum size=0.5cm, rounded corners=6pt] at (2.5,-1.5) {$\sigma(.)$};

        \node (output1) at (5.3,0.5) {\(\bbX^{(L)}\)};

        \draw[->] (x0) -- (DCN1);  
        \draw[->] (DCN1) -- (dots1);
        \draw[->] (dots1) to[out=0, in=180] (DCNl1);
        \draw[->] (DCNl1) to[out=0, in=180] (dots2);
        \draw[->] (dots2) to[out=0, in=180] (DCNL);
        \draw[->] (DCNL) -- (output1);
        \draw[->] (DCNl) -- (sig);
        \draw[->] (xl2) -- (DCNl);
        \draw[->] (sig) -- (xl3);

        \draw[-] (DCNl1) to[out=-40, in=160] (B);
        \draw[-] (DCNl1) to[out=220, in=20] (A);

        \draw[dashed, thick, rounded corners=15pt] (-3.5,2.5) rectangle (4.6,-2);

        \end{tikzpicture}
        \subcaption{DCN architecture}
    \end{minipage}%
    \hfill
    \begin{minipage}{0.48\textwidth}
        \centering
        \begin{tikzpicture}[
            scale=0.9,
            ->, >=stealth',
            node distance=1.5cm and 1.8cm,
            every node/.style={font=\small},
            main/.style={rectangle, draw, fill=gray!10, minimum size=0.8cm, text centered},
            source/.style={rectangle, draw, fill=gray!30, minimum size=0.8cm, text centered}
        ]
        
        \node (x) at (6.5,0.5) {$\bbX$};
        
        \node (T1) at (9,2) [draw, fill=gray!10, rounded corners=3pt, minimum width=1.8cm, minimum height=0.6cm] {$\bbS_1 \bbX$};
        \node (T2) at (9,0.5) [draw, fill=gray!10, rounded corners=3pt, minimum width=1.8cm, minimum height=0.6cm] {$\bbS_2 \bbX$};
        \node (TN) at (9,-1.5) [draw, fill=gray!10, rounded corners=3pt, minimum width=1.8cm, minimum height=0.6cm] {$\bbS_N \bbX$};
        
        \node (MLP1) at (12,2) [draw, fill=gray!30, rounded corners=3pt, minimum width=1.4cm, minimum height=0.6cm] {$\mlp$};
        \node (MLP2) at (12,0.5) [draw, fill=gray!30, rounded corners=3pt, minimum width=1.4cm, minimum height=0.6cm] {$\mlp$};
        \node (MLPN) at (12,-1.5) [draw, fill=gray!30, rounded corners=3pt, minimum width=1.4cm, minimum height=0.6cm] {$\mlp$};
        
        \node at (9,-0.45) {\vdots};
        \node at (11.9,-0.45) {\vdots};
        
        \node (sum) at (14,0.5) [circle, draw, minimum size=0.6cm] {\(\sum\)};
        \node (output) at (15.5,0.5) {$\bbZ$};
        
        \draw[->, thick] (x.east) to[out=20, in=160] (T1.west);
        \draw[->, thick] (x.east) to[out=0, in=180] (T2.west);
        \draw[->, thick] (x.east) to[out=-20, in=200] (TN.west);
        
        \draw[->, thick] (T1.east) -- (MLP1.west);
        \draw[->, thick] (T2.east) -- (MLP2.west);
        \draw[->, thick] (TN.east) -- (MLPN.west);
        
        \draw[->, thick] (MLP1.east) to[out=0, in=150] (sum.west);
        \draw[->, thick] (MLP2.east) to[out=0, in=180] (sum.west);
        \draw[->, thick] (MLPN.east) to[out=0, in=210] (sum.west);
        
        \draw[->, thick] (sum.east) -- (output.west);
        
        \draw[dashed, thick, rounded corners=15pt] (6.92,2.5) rectangle (14.6,-2);

        \end{tikzpicture}
        \subcaption{PDCN architecture}
    \end{minipage}
    \caption{Comparison of the DCN (a) and PDCN (b) architectures. The DCN is structured sequentially as a deep architecture, processing features layer by layer. On the other, hand the PDCN processes multiple shifted signals $\bbS_k \bbX$ in parallel across separate branches, resulting in a wider architecture. When the learnable weights of the MLP are shared, the number of PDCN parameters is independent of the number of nodes $N$.}
    \label{fig:DCN_PDCN}
\end{figure*}
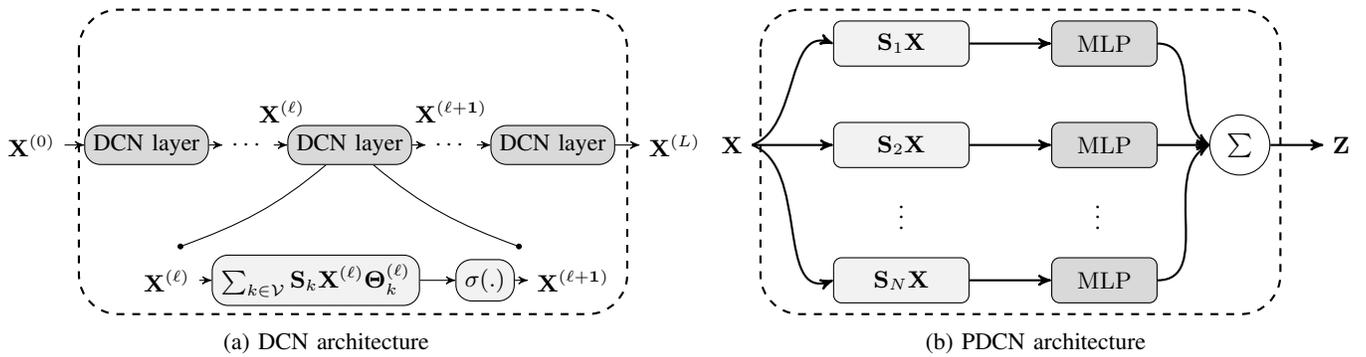

%
%

\section{DAG Convolutional Network (DCN)}\label{ssec:DCN}
We now have all the ingredients needed to introduce a convolutional network for signals defined on DAGs, henceforth referred to as DCN.
Central to this end is the definition of causal graph filters in \eqref{eq:causal_gf}, which we use to implement layerwise convolutions and compose them with a pointwise nonlinearity to obtain a DAG perceptron.  
The DCN is a deep learning architecture obtained by stacking multiple such DAG perceptrons. In its simplest form, the output of the layer $\ell = 1, \ldots, L$ is recursively given by
\begin{equation}\label{eq:simple_dcn}
    \bbx^{(\ell)} = \sigma \left( \sum_{k \in \ccalV} \theta_k^{(\ell)} \bbS_k \bbx^{(\ell-1)} \right),
\end{equation}
where $\theta_k^{(\ell)}$ denote the learnable filter coefficients, and $\sigma(\cdot)$ is a nonlinear activation function, such as the $\mathrm{ReLU}$. The input $\bbx^{(0)}=\bbx$ to the first layer is the graph signal $\bbx$ we want to process using the DCN.
The output of the final layer $\bbx^{(L)}$ is a graph signal collecting (here scalar, or single feature) nodal representations that can be used for different downstream tasks; see Section~\ref{sec:experiments} for several examples. Extensions to graph signals with multiple features are straightforward, and will be addressed later in this section. 

The recursion \eqref{eq:simple_dcn} can be interpreted in two complementary ways.
From a spectral perspective, the operation  $\bbS_k \mathbf{x}^{(\ell-1)} = \bbW \bbD_k \bbc^{(\ell-1)}$ first selects the \emph{causes} (i.e., the exogenous inputs in an SEM) of predecessors of node {$ v_k $}.
These causes are then diffused across the reachabilty DAG $\bar{\ccalD}$ via the weighted transitive closure matrix $\bbW$, effectively infusing the inductive bias stemming from the partial ordering of vertices in the DAG.
Alternatively, this operation can be understood from the perspective of message-passing GNNs.
Specifically, for each node $ {v_i} \in \ccalV $, every GSO $ \bbS_k $ aggregates features from common predecessors of nodes {$ v_i $} and {$ v_k $}, creating different messages incoming at {$v_i$}.
{These messages are then linearly combined and
passed through a point-wise nonlinear activation $\sigma(\cdot)$ to update the node representations $\bbx^{(\ell)}$ at the output of layer $\ell$.
}


Although the single-filter DCN in \eqref{eq:simple_dcn} provides an intuitive foundation to understand the DCN model, it may lack sufficient expressiveness to learn complex DAG signal representations.
Moreover, since input signals often have multiple features ($F>1$), a more flexible architecture is required.
To address these shortcomings, we extend the single-filter formulation in \eqref{eq:simple_dcn} using a bank of learnable DAG filters, leading to [cf. \eqref{eq:fb_gnn}]
\begin{equation}\label{eq:dcn_layer}
    \mathbf{X}^{(\ell)} = \sigma \left( \sum_{k \in \ccalV} \bbS_k \mathbf{X}^{(\ell-1)} \bbTheta_k^{(\ell)} \right).
\end{equation}
Here, $\bbTheta_k^{(\ell)} \in \mathbb{R}^{F_i^{(\ell)} \times F_o^{(\ell)}}$ contains the learnable filter coefficients, where $F_i^{(\ell)}$ and $F_o^{(\ell)}$ respectively denote the number of input and output features. Naturally, $F_i^{(\ell)}=F_o^{(\ell-1)}$.  Fig.~\ref{fig:DCN_PDCN}a depicts this recursion for a DCN with $L$ layers, where the input is $\bbX^{(0)}=\bbX$ and the output is given by $\mathbf{X}^{(L)} = f_{\bbTheta}(\mathbf{X};\ccalS)$, with $\ccalS = \{ \bbS_k \}_{k \in \ccalV}$ being the set of causal GSOs and $\bbTheta$ collects all the learnable parameters. The number of layers $L$ and features $F_i^{(\ell)},F_o^{(\ell)}$ are hyperparameters chosen before training. Given a training set $ \ccalT = \{(\bbX_m, \bby_m)\}_{m=1}^M $, learning is accomplished by using mini-batch stochastic gradient descent to minimize the task-dependent empirical risk \eqref{eq:emp_risk_min}.

While the bank of learnable filters enhances the expressiveness of DCN, feature aggregation remains dictated by the causal GSOs, ensuring the model exploits the structural information of the DAG.
In fact, the attentive reader will have noticed our slight abuse of notation when using $f_{\bbTheta}(\cdot ; \ccalS)$ instead of $f_{\bbTheta}(\cdot ; \ccalD)$ as introduced in Section~\ref{ssec:prob_statement}.
Formally these are not the same, since the causal shifts in $\ccalS$ are not uniquely defined given $\ccalD$ (recall there is a choice to be made regarding the weighted transitive closure).
In any case, what we want to convey is that the prior structural information in $\ccalD$ enters the DCN model via $\ccalS$. 
Next, we elaborate on the advantages and limitations of the DCN.

%
%

\subsection{Analysis and Discussion}
The DCN convolutional layer definition in \eqref{eq:dcn_layer} resembles the GNN architectures designed for general (cyclic) graphs; see also the GNN recursion in (2).
However, relying on causal graph filters is essential to implement formal convolutions over DAGs, and this
innovation offers several advantages.
First, the spectrum of $\bbS_k$  is well defined. Endowing the DCN with a spectral representation could pave the way to studying properties such as stability, transferability, or denoising capability~\cite{ruiz2021graph,rey2022untrained}.
Exploring these properties further presents a promising direction for future research beyond the scope of this paper.
In addition, the eigenvalues of $ \bbS_k $ are the diagonal entries of the binary matrices $ \bbD_k $.
When combined with non-expansive activation functions such as the $\mathrm{ReLU}$, this ensures numerical stability even when multiple layers are stacked, avoiding issues that commonly arise in other GNN architectures{\cite{peng2024beyond,arroyo2025vanishing}}.
In other words, causal GSOs are inherently well conditioned and there is no need for e.g., degree-based normalization of aggregation operators to keep the eigenvalue spread in check.

From a computational perspective, alternative architectures designed to learn from DAG-related data involve \emph{aggregation} and \emph{combination} operations that are computationally demanding~\cite{zhang2019dvaevariationalautoencoderdirected,thost2021directed}.
In contrast, in DCN the \emph{aggregation} is determined by the causal GSOs via the products $\bbS_k\bbX^{(\ell-1)}$, and the matrices of learnable coefficients $\bbTheta_k^{(\ell)}$ determine the weights of the \emph{combination} step. 
{Here, the weighted transitive closure $\bbW$ (and thus the causal GSOs) can be computed once offline prior to training.
Furthermore, recall that the diagonal matrices $\bbD_k$ have binary entries (hence $\bbD_k=\bbD_k^2$).
As a result, the corresponding causal GSOs are idempotent and contain all-zero columns, leading to numerically stable and efficient layer-wise transformations.}
Fig.~\ref{fig:example} shows that causal GSOs are typically quite sparse.
{Hence, the DCN layers in \eqref{eq:dcn_layer} only involve sparse matrix products.
More formally, let $S=\max_{k\in\ccalV}\|\bbS_k\|_0$ denote the number of nonzero entries of the densest causal GSO, with $S \ll N^2$.
Assuming a fixed number of features $F$ for simplicity, the worst-case computational complexity of a DCN layer is $\ccalO\!\left(NF(S+NF)\right)$.
Compared with classical convolutional GNNs, this complexity includes an additional factor of $N$, which stems from the use of $N$ distinct causal GSOs and will be discussed shortly.}
In contrast, related architectures such as D-VAE~\cite{zhang2019dvaevariationalautoencoderdirected} or DAGNN~\cite{thost2021directed} require more complex and computationally demanding sequential operations, resulting in longer running times. The number of sequential propagation steps in DAGNN is determined by the depth of the input DAG, i.e., by the length of its longest source-to-sink path. Nodes within the same topological level are updated in parallel, so wide but shallow graphs require only a few iterations regardless of their size, whereas long, chain-like graphs represent the worst case. In batched processing, the number of iterations is determined by the deepest graph in the batch; see also the numerical tests in Section~\ref{sec:experiments} and~\cite[Theorem~1]{thost2021directed}.


The labeling of the nodes in $\ccalV$ is in principle arbitrary.
Thus, to ensure model consistency, the output of a DCN should be robust to node relabelings.
Building on the permutation equivariance property of causal graph filters in Theorem~\ref{thm:gf_equivariant}, the following corollary establishes that DCNs are permutation equivariant mappings as well.
\begin{corollary}\label{col:permutation_eq_dcn}
Let $\ccalV'$ be the poset of nodes obtained by permuting with $\pi:\ccalV\mapsto \ccalV'$ the DAG with node ordering $\ccalV$. Let $\bbP$ be the matrix representation of $\pi$.
The DCN output $f_{\bbTheta}(\bbX ; \ccalS)$ is permutation equivariant, i.e.,
\begin{equation}
        f_{\bbTheta'}(\bbX' ; \ccalS') = \bbP f_{\bbTheta}(\bbX ; \ccalS),
\end{equation}
where $\ccalS = \{ \bbS_k\}_{k \in \ccalV}$ is the set collecting all causal GSOs, $\bbX'=\bbP\bbX$ are the permuted inputs, and $\ccalS' = \{\bbS_q'\}_{q \in \ccalV'}$ and $\bbTheta' = \{\bbTheta_q\}_{q \in \ccalV'}$ are the permuted causal GSOs and learnable parameters, respectively.
\end{corollary}
\begin{proof}
We prove the result for a DAG perceptron for simplicity, though the argument extends to multiple layers by induction.
Leveraging Theorem~\ref{thm:gf_equivariant}, from \eqref{eq:dcn_layer} we obtain
\begin{align}
        f_{\bbTheta'}(\bbX' ; \ccalS') &= \sigma\left(\sum_{q \in \ccalV'} \bbS_q' \bbX' \bbTheta_q\right) \nonumber \\
        &= \sigma\left(\bbP \sum_{k \in \ccalV} \bbS_k \bbX \bbTheta_k\right) 
        = \bbP f_{\bbTheta}(\bbX ; \ccalS),
\end{align}
where, as in the proof of Theorem~\ref{thm:gf_equivariant}, $\bbTheta_k$ denotes the matrix of learnable coefficients associated with {$v_k=\pi^{-1}(v_q)$}. 
{The last equality holds because $\sigma(\cdot)$ acts elementwise and left-multiplication by $\bbP$ permutes the rows of the rightmost matrix factor, hence $\sigma(\bbP\bbZ)=\bbP\sigma(\bbZ)$, for any $\bbZ$.
}
\end{proof}

On the downside, the number of GSOs involved in the convolution, and consequently the number of learnable parameters, increases linearly with the size of the graph $N$.  
This scaling poses challenges in terms of computational cost and memory requirements.  
An engineering workaround to mitigate this issue is to approximate the convolution in \eqref{eq:dcn_layer} by restricting the summands therein to a subset of nodes $\ccalU \subset \ccalV$ with fixed cardinality, independent of $N$.
Admittedly, this reduction still leads to a convolutional filter but it may limit the expressiveness of the DCN.  
However, this limitation can be alleviated by stacking multiple layers, as the cross-products of different $\bbS_k$ and $\bbS_\ell$ can give rise to a new GSO. Furthermore, when the number of selected GSOs satisfies $|\ccalU| \ll N$, the computational complexity of the DCN reduces to $\ccalO (F(S+NF))$. This is comparable to GNNs for general graphs with cycles. In this work, we observed that selecting the GSOs involved in the convolution via uniform sampling yields competitive performance; the numerical experiments in Section~\ref{sec:experiments} offer evidence supporting this claim. Nevertheless, developing a principled method to determine the subset of nodes in $\ccalU$ remains a promising direction for future research.
Finally, note that Corollary~\ref{col:permutation_eq_dcn} assumes the use of all nodes in $\ccalV$. However, the permutation equivariance of DCN using only GSOs related to a subset of nodes $\ccalU$ still holds, provided that the permutation $\pi$ only acts on nodes within this set.

%
%

\section{Parallel DCN}\label{ssec:PDCN}
The DCN offers a principled framework for learning representations from data on DAGs. Because causal GSOs are idempotent operators~\cite{seifert2023causal}, the benefits of employing a deep architecture to gain expressive power are less clear, at least when using all of the GSOs. In such a case, and disregarding the (important) effect of interleaving pointwise nonlinearities, one can show that the composition of two causal graph filters can always be re-parameterized using a single causal graph filter. Building on this observation, we introduce Parallel DCN (PDCN), a variant of DCN that shifts from the sequential structure in \eqref{eq:dcn_layer} to a parallel (wide) implementation. We show that the number of PDCN parameters is independent of $N$.

The proposed PDCN architecture comprises $N$ parallel branches, with shared input $\bbX$ and respective outputs
\begin{equation}\label{eq:pdcn_branch}
    \barbZ_k = \mathrm{MLP} \left( \bbS_k \bbX\right) = \sigma \left(\dots \sigma\left( \bbS_k \bbX \bbTheta^{(1)}\right)\dots \bbTheta^{(L)} \right),
\end{equation}
where $\bbTheta^{(\ell)}$ now denotes the learnable parameters of the $\ell$-th MLP layer. Notably, these weights are common to all branches. Inspired by~\cite{doshi2022graph}, the MLP in each branch {$v_k\in\ccalV$} is fed the shifted signal $\bbS_k\bbX$. The PDCN output is computed by aggregating the output of every branch, namely
\begin{equation}\label{eq:pdcn_readout}
    \bbZ = \sum_{k \in \ccalV} \barbZ_k.
\end{equation}
%

Similarly to the DCN layers in \eqref{eq:dcn_layer}, PDCN also exploits the causal GSOs to aggregate information from the ancestors of each node in a principled way.
These DAG-aware features are then fed to an MLP, which learns how to combine them. The PDCN architecture is depicted in Fig.~\ref{fig:DCN_PDCN}b. Observe how, while DCN is a sequential architecture amenable to grow deeper, PDCN exhibits a parallel structure with tunable width (say by dropping a subset of causal GSOs indexed by $\ccalV\setminus \ccalU$). Another benefit of this alternative design is that PDCN is amenable to parallelization, leading to efficient inference.

%
%

\subsection{Analysis and Discussion}
If the primary concern is model expressivity, each PDCN branch could employ a different MLP with parameters $\bbTheta_k^{(\ell)}$, {$v_k\in\ccalV$}.  
{However, we advocate for a weight-sharing strategy across all branches.}
Like a siamese network, all MLPs in PDCN share the same set of weights $\bbTheta^{(\ell)}$, effectively decoupling the number of learnable parameters from the size of the graph. Assuming a fixed number of features $F$, this approximately adds up to a total of $\ccalO (F^2L)$ learnable parameters.
{The upshot is a lighter architecture that helps mitigate overfitting and reduces the computational cost.}

Even with weight sharing, the computational complexity scales linearly with $N$. Following the same reasoning as in DCN, one can trade-off expressivity for complexity by using only a subset of the available GSOs, i.e., selecting $|\ccalU| \leq N$ branches.
{Moreover, in the specific case of PDCN, the shifted signals $\bbS_k \bbX$ can be precomputed prior to training, further alleviating the computational burden. As a result, the runtime complexity is exclusively determined by the width and depth of the MLP and by the number of active branches.}
Weight sharing also offers the added flexibility of training a PDCN model using only some of the branches, and then performing inference with other, more, or all GSOs enabled as desired. 

Once more, performing the aggregation step using causal GSOs ensures that PDCN is permutation equivariant.  
\begin{corollary}\label{col:permutation_eq_pdcn}
Let $\ccalV'$ be the poset of nodes obtained by permuting with $\pi:\ccalV\mapsto \ccalV'$ the DAG with node ordering $\ccalV$. Let $\bbP$ be the matrix representation of $\pi$. The PDCN output $f^p_{\bbTheta}(\bbX ; \ccalS)$ is permutation equivariant, i.e.,
\begin{equation}
    f^p_{\bbTheta}(\bbX' ; \ccalS') = \bbP f^p_{\bbTheta}(\bbX ; \ccalS),
\end{equation}
where $\ccalS = \{ \bbS_k\}_{k \in \ccalV}$ is the set collecting all causal GSOs, $\bbX'=\bbP\bbX$ are the permuted inputs, and $\ccalS' = \{\bbS_q'\}_{q \in \ccalV'}$ are the permuted causal GSOs.
\end{corollary}
\begin{proof}
From Theorem~\ref{thm:gf_equivariant} we have $\bbS_q' = \bbP \bbS_k \bbP^\top$, for ${v_q=\pi(v_k)}$. Given the input $\bbX'$, the output of each branch is given by  
\begin{equation}
\bbZ_q' = \sigma \left( \bbS_q' \bbX' \bbTheta^{(1)}\right) = \bbP \sigma \left( \bbS_k \bbX \bbTheta^{(1)} \right) = \bbP \bbZ_k.
\end{equation}
Here, we consider a single-layer MLP for simplicity, but the same reasoning applies to an MLP with $L$ layers. Finally, since 
\begin{equation}
f^p_{\bbTheta}(\bbX' ; \ccalS') = \sum_{q \in \ccalV'} \bbZ_q' = \bbP \sum_{k \in \ccalV} \bbZ_k = \bbP f^p_{\bbTheta}(\bbX ; \ccalS),
\end{equation}
the result follows.  
\end{proof}
Corollary~\ref{col:permutation_eq_pdcn} assumes a PDCN with $N$ branches, but the result still holds for a smaller number of branches as long as the permutation affects only those GSOs in the model.

Finally, an important property of GNNs is their ability to distinguish between non-isomorphic graphs, as characterized by the WL test~\cite{WL1968test}. 
{In what follows, we argue that in light of previous results for parallel aggregation architectures, PDCN satisfies this property for non-isomorphic DAGs under standard injectivity assumptions.}
Parallel GNNs distinguish non-isomorphic graphs if their parallel aggregation and readout functions are injective~\cite[Th. 1]{doshi2022graph}. PDCN belongs to this class of parallel GNNs, with its aggregation and readout functions defined in \eqref{eq:pdcn_branch} and \eqref{eq:pdcn_readout}, respectively.  
From a node-wise perspective, this diffusion can be expressed as  
\begin{equation}
    \sum_{v \leq u} [ \bbS_k ]_{uv} \bbx_v,\quad {v_u}
    \in\ccalV,
\end{equation}
where each node $u$ aggregates the features $\bbx_v$ from its predecessors {$v_v\leq v_u$} (i.e., its in-neighbors in the reachability DAG $\bar{\ccalD}$).
{Proposition~\ref{prop:uniqueness} establishes that the set of causal GSOs $\ccalS$ is uniquely determined by the underlying DAG. Hence, for two non-isomorphic DAGs, there exists at least one branch for which the corresponding aggregation operator differs.}
{Moreover, under one-hot encoded node features, neighborhood aggregation via summation is injective over multisets.} 
In particular,~\cite[Lemma 5]{xu2019powerfulgraphneuralnetworks} asserts that there exists a function $\phi$ such that $\sum_{x \in \ccalX} \phi(x)$ is unique for each multiset $\ccalX$.
{Therefore, by the universal approximation theorem and under sufficient model capacity, the MLP in PDCN can approximate injective mappings and thereby yield an injective global readout.}
\begin{remark}\normalfont
Based on the preceding discussion, {and under the injectivity conditions outlined above}, PDCN satisfies the assumptions of~\cite[Th. 1]{doshi2022graph} and is therefore capable of mapping non-isomorphic DAGs, as recognized by the WL test, to distinct embeddings.
\end{remark}

\section{Numerical Evaluation}\label{sec:experiments}
We demonstrate the strong performance of the DCN architecture across diverse tasks and settings, through comprehensive benchmarking against state-of-the-art models using both synthetic and real-world datasets.
{We compare against the DAG-agnostic GNN baselines GCN~\cite{kipf2016semi},  FB-GCNN~\cite{ruiz2021graph}, GraphSAGE~\cite{hamilton2018inductiverepresentationlearninglarge}, GIN~\cite{xu2019powerfulgraphneuralnetworks}, GAT~\cite{petar2018graphattentionnetworks}, GGNN~\cite{li2016gated}, and Scarselli-GNN~\cite{scarselli2008graph}.
We also consider MLP~\cite{rumelhart1986learning} and the sequential models LSTM~\cite{lstm}, GRU~\cite{gru} to probe the added value of leveraging the graph structure.
In terms of DAG-aware baselines, we consider D-VAE~\cite{zhang2019dvaevariationalautoencoderdirected}, DAGNN~\cite{thost2021directed}, 
and~DAG+NodeFormer~\cite{luo2023transformers}. 
Note that our GSP focus is on learning from DAG signals, rather than learning a latent representation for DAGs.}
See Appendix \ref{app:imp_details} for additional details on the implementation of these baselines.

Synthetic experiments in Section \ref{ssec:synthetic_data} include two tasks: prediction of a diffused signal and source identification. In the diffusion learning task, we evaluate the ability of the model to capture signal propagation dynamics over DAGs. After training the graph signal regression model using an MSE loss $\ccalL_{\textrm{MSE}}$, performance is assessed using the normalized MSE (NMSE), computed as:
$\textrm{NMSE}:=\frac{1}{T} \sum_{t=1}^T \frac{\| \bby_t - \hat{\bby}_t \|_2^2}{\| \bby_t \|_2^2}$
where $T$ denotes the number of test signals, $\hat{\bby}_t$ is the predicted graph signal and $\bby_t$ the corresponding ground-truth response. For the source identification task, the goal is to evaluate the model's ability to trace the source of a signal after observing its propagation through downstream nodes. We treat this as a node classification problem, where the true source is selected from a subset of unobserved upstream nodes in the DAG. The cross-entropy loss $\ccalL_{\textrm{CE}}$ is employed during training, and the performance is reported in terms of classification accuracy. 
Additionally, we conduct experiments on two real-world datasets: {a gene expression dataset and a hydrological dataset from the River Thames.}
In all cases, the data is split into 70\% for training, 20\% for validation, and 10\% for testing. The code to reproduce all results, along with additional experiments and implementation details, is available on GitHub\footnote{\href{https://github.com/reysam93/dag_conv_nn/}{\small\texttt{https://github.com/reysam93/{dag\_conv\_nn}/}}}.

\begin{table}[t]
    \centering
    \resizebox{\columnwidth}{!}{
\begin{tabular}{l|c c|c c}
    & \multicolumn{2}{c|}{\textbf{Diffusion Learning}} 
    & \multicolumn{2}{c}{\textbf{Source Identification}} \\
    & NMSE & Time (s) & Accuracy & Time (s) \\
    \hline
    DCN & $\mathbf{0.014 \pm 0.010}$ & 7.9 & $0.057 \pm 0.008$ & 22.2 \\
    DCN-30 & $0.029 \pm 0.017$ & 6.0 & $0.053 \pm 0.012$ & 20.1 \\
    DCN-10 & $0.049 \pm 0.021$ & 5.9 & $0.054 \pm 0.005$ & 17.0 \\
    \hline
    DCN-T & $0.073 \pm 0.024$ & 8.1 & $\mathbf{0.997 \pm 0.018}$ & 22.5 \\
    DCN-30-T & $0.153 \pm 0.030$ & 6.0 & $0.996 \pm 0.032$ & 17.8 \\
    DCN-10-T & $0.181 \pm 0.030$ & 5.9 & $0.933 \pm 0.141$ & 19.5 \\
    \hline
    DAGNN & $0.173 \pm 0.040$ & 5111.8 & $0.957 \pm 0.031$ & 9344.5 \\
    D-VAE & $0.163 \pm 0.041$ & 14024.1 & $0.987 \pm 0.007$ & 12841.9 \\
    {DAG+NodeFormer} 
        & {$0.581 \pm 0.016$} & {744.1}
        & {$0.940 \pm 0.068$} & {2725.3} \\
    \hline
    LS & $0.048 \pm 0.022$ & 0.4 & $0.06 \pm 0.016$ & 0.45 \\
    PDCN & $0.098 \pm 0.015$ & 22.6 & $0.288 \pm 0.114$ & 79.5 \\
    FB-GCNN & $0.138 \pm 0.028$ & 5.8 & $0.676 \pm 0.172$ & 18.6\\
    GraphSAGE & $0.341 \pm 0.039$ & 9.9 & $0.620 \pm 0.163$ & 40.0 \\
    GIN & $0.404 \pm 0.079$ & 9.7 & $0.176 \pm 0.098$ & 39.8 \\
    GCN & $0.167 \pm 0.037$ & 5.2 & $0.052 \pm 0.018$ & 7.6 \\
    GAT & $0.653 \pm 0.089$ & 19.9 & $0.071 \pm 0.018$ & 38.5\\
    MLP & $0.392 \pm 0.012$ & 4.1 & $0.050 \pm 0.016$ & 16.9 \\
    {Scarselli-GNN}
        & {$0.033 \pm 0.006$} & {172.5}
        & {$0.185 \pm 0.132$} & {127.2} \\
    {GRU}
        & {$0.304 \pm 0.035$} & {13.3}
        & {$0.324 \pm 0.025$} & {14.8} \\
    {LSTM}
        & {$0.296 \pm 0.043$} & {13.3}
        & {$0.311 \pm 0.034$} & {16.1} \\
    {GGNN}
        & {$0.530 \pm 0.019$} & {405.0}
        & {$0.778 \pm 0.351$} & {907.0} \\
    \hline
\end{tabular}
    }
    \caption{NMSE and accuracy for the network diffusion learning and source identification tasks, respectively. The mean performance, standard deviation and average training plus inference time across 25 independent realizations are reported.}
    \label{t:table1}
\end{table}

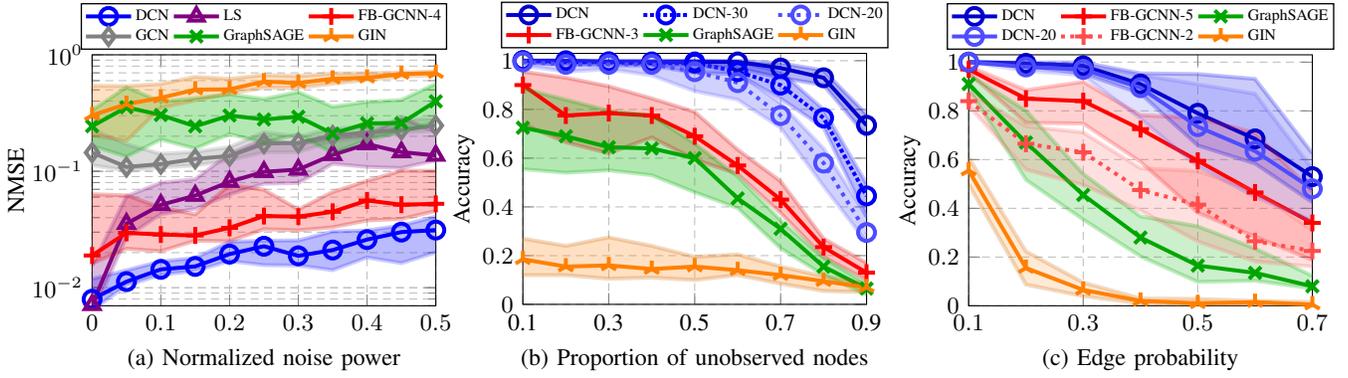
\begin{figure*}[!t]
	\centering
	\begin{subfigure}{0.32\textwidth}
		\centering
		 \begin{tikzpicture}[baseline,scale=1]

\pgfplotstableread{data/noise/noise_inf-constant-med_err.csv}\errtable
\pgfplotstableread{data/noise/noise_inf-constant-prct75_err.csv}\prcttop
\pgfplotstableread{data/noise/noise_inf-constant-prct25_err.csv}\prctbot

\pgfmathsetmacro{\opacity}{0.3}
\pgfmathsetmacro{\contourop}{0.25}

\begin{semilogyaxis}[
    xlabel={(a) Normalized noise power},
    xmin=0,
    xmax=0.5,
    xtick = {0, .1, ..., .51},
    ylabel={NMSE},
    ymin = .007,
    ymax = 1,
    grid style=densely dashed,
    grid=both,
    legend style={
        at={(.5, 1.02)},
        anchor=south},
    legend columns=3,
    width=175,
    height=165,
    ]

    \addplot [blue!80!white, name path = DCN-bot, opacity=\contourop, forget plot] table [x=xaxis, y=DCN] \prctbot;
    \addplot [blue!90!white, name path = DCN-top, opacity=\contourop, forget plot] table [x=xaxis, y=DCN] \prcttop;
    \addplot[blue!90!white, fill opacity=\opacity, forget plot] fill between[of=DCN-bot and DCN-top];
    \addplot[blue, mark=o, solid] table [x=xaxis, y=DCN] {\errtable};

    \addplot [violet!90!white,name path = Linear-bot, opacity=\contourop, forget plot] table [x=xaxis, y=Linear] \prctbot;
    \addplot [violet!90!white,name path = Linear-top, opacity=\contourop, forget plot] table [x=xaxis, y=Linear] \prcttop;
    \addplot[violet!90!white, fill opacity=\opacity, forget plot, forget plot] fill between[of=Linear-bot and Linear-top];
    \addplot[violet, mark=triangle , solid] table [x=xaxis, y=Linear] {\errtable};
    
    \addplot [red!80!black, name path = FB-GCNN-bot, opacity=\contourop, forget plot] table [x=xaxis, y=FB-GCNN-4] \prctbot;
    \addplot [red!80!black, name path = FB-GCNN-top, opacity=\contourop, forget plot] table [x=xaxis, y=FB-GCNN-4] \prcttop;
    \addplot[red!80!white, fill opacity=\opacity, forget plot, forget plot] fill between[of=FB-GCNN-bot and FB-GCNN-top];
    \addplot[red, mark=+, solid] table [x=xaxis, y=FB-GCNN-4] {\errtable};

    \addplot [gray!90!white,name path = GNN-bot, opacity=\contourop, forget plot] table [x=xaxis, y=GNN-A] \prctbot;
    \addplot [gray!90!white,name path = GNN-top, opacity=\contourop, forget plot] table [x=xaxis, y=GNN-A] \prcttop;
    \addplot[gray!90!white, fill opacity=\opacity, forget plot, forget plot] fill between[of=GNN-bot and GNN-top];
    \addplot[gray, mark=diamond , solid] table [x=xaxis, y=GNN-A] {\errtable};

    \addplot [green!70!black,name path = GraphSAGE-bot, opacity=\contourop, forget plot] table [x=xaxis, y=GraphSAGE-A] \prctbot;
    \addplot [green!70!black,name path = GraphSAGE-top, opacity=\contourop, forget plot] table [x=xaxis, y=GraphSAGE-A] \prcttop;
    \addplot[green!70!black, fill opacity=\opacity, forget plot, forget plot] fill between[of=GraphSAGE-bot and GraphSAGE-top];
    \addplot[green!65!black, mark=x, solid] table [x=xaxis, y=GraphSAGE-A] {\errtable};
    
    \addplot [orange!80!black, name path = GIN-bot, opacity=\contourop, forget plot] table [x=xaxis, y=GIN-A] \prctbot;
    \addplot [orange!80!black, name path = GIN-top, opacity=\contourop, forget plot] table [x=xaxis, y=GIN-A] \prcttop;
    \addplot[orange!80!white, fill opacity=\opacity, forget plot, forget plot] fill between[of=GIN-bot and GIN-top];
    \addplot[orange, mark=Mercedes star, solid] table [x=xaxis, y=GIN-A] {\errtable};

    \legend{DCN, LS, FB-GCNN-4, GCN, GraphSAGE, GIN}
\end{semilogyaxis}
\end{tikzpicture}\label{fig:exp_a}
	\end{subfigure}
	\begin{subfigure}{0.32\textwidth}
		\centering
        \begin{tikzpicture}[baseline,scale=1]

\pgfplotstableread{data/src_nodes/src_nodes_inf-random-med_acc.csv}\acctable
\pgfplotstableread{data/src_nodes/src_nodes_inf-random-prct75_acc.csv}\prcttop
\pgfplotstableread{data/src_nodes/src_nodes_inf-random-prct25_acc.csv}\prctbot

\pgfmathsetmacro{\opacity}{0.3}
\pgfmathsetmacro{\contourop}{0.25}

\begin{axis}[
    xlabel={(b) Proportion of unobserved nodes},
    xmin=10,
    xmax=90,
    xtick = {10, 30, ..., 90},
    xticklabels = {0.1, 0.3, 0.5, 0.7, 0.9},
    ylabel={Accuracy},
    ymin = 0,
    ymax = 1.03,
    ytick = {0, .2, ..., 1},
    grid style=densely dashed,
    grid=both,
    legend style={
        at={(.5, 1.02)},
        anchor=south},
    legend columns=3,
    width=175,
    height=165,
    ]

    \addplot [blue!90!black, name path = DCN-bot, opacity=\contourop, forget plot] table [x=xaxis, y=DCN] \prctbot;
    \addplot [blue!90!black, name path = DCN-top, opacity=\contourop, forget plot] table [x=xaxis, y=DCN] \prcttop;
    \addplot[blue!90!black, fill opacity=\opacity, forget plot] fill between[of=DCN-bot and DCN-top];
    \addplot[blue!80!black, mark=o, solid] table [x=xaxis, y=DCN] {\acctable};
    
    \addplot [blue!90!white,name path = DCN-30-bot, opacity=\contourop, forget plot] table [x=xaxis, y=DCN-30] \prctbot;
    \addplot [blue!90!white,name path = DCN-30-top, opacity=\contourop, forget plot] table [x=xaxis, y=DCN-30] \prcttop;
    \addplot[blue!90!white, fill opacity=\opacity, forget plot] fill between[of=DCN-30-bot and DCN-30-top];
    \addplot[blue, mark=o, densely dotted] table [x=xaxis, y=DCN-30] {\acctable};

    \addplot [blue!60!white, name path = DCN-20-bot, opacity=\contourop, forget plot] table [x=xaxis, y=DCN-20] \prctbot;
    \addplot [blue!60!white, name path = DCN-20-top, opacity=\contourop, forget plot] table [x=xaxis, y=DCN-20] \prcttop;
    \addplot[blue!60!white, fill opacity=\opacity, forget plot] fill between[of=DCN-20-bot and DCN-20-top];
    \addplot[blue!70!white, mark=o, dotted] table [x=xaxis, y=DCN-20] {\acctable};


    \addplot [red!80!black, name path = FB-GCNN-bot, opacity=\contourop, forget plot] table [x=xaxis, y=FB-GCNN-3] \prctbot;
    \addplot [red!80!black, name path = FB-GCNN-top, opacity=\contourop, forget plot] table [x=xaxis, y=FB-GCNN-3] \prcttop;
    \addplot[red!80!white, fill opacity=\opacity, forget plot, forget plot] fill between[of=FB-GCNN-bot and FB-GCNN-top];
    \addplot[red, mark=+, solid] table [x=xaxis, y=FB-GCNN-3] {\acctable};

    \addplot [green!70!black,name path = GraphSAGE-bot, opacity=\contourop, forget plot] table [x=xaxis, y=GraphSAGE-A] \prctbot;
    \addplot [green!70!black,name path = GraphSAGE-top, opacity=\contourop, forget plot] table [x=xaxis, y=GraphSAGE-A] \prcttop;
    \addplot[green!70!black, fill opacity=\opacity, forget plot, forget plot] fill between[of=GraphSAGE-bot and GraphSAGE-top];
    \addplot[green!65!black, mark=x, solid] table [x=xaxis, y=GraphSAGE-A] {\acctable};

    \addplot [orange!80!black, name path = GIN-bot, opacity=\contourop, forget plot] table [x=xaxis, y=GIN-A] \prctbot;
    \addplot [orange!80!black, name path = GIN-top, opacity=\contourop, forget plot] table [x=xaxis, y=GIN-A] \prcttop;
    \addplot[orange!80!white, fill opacity=\opacity, forget plot, forget plot] fill between[of=GIN-bot and GIN-top];
    \addplot[orange, mark=Mercedes star, solid] table [x=xaxis, y=GIN-A] {\acctable};

    \legend{DCN, DCN-30, DCN-20, FB-GCNN-3, GraphSAGE, GIN}
\end{axis}
\end{tikzpicture}
	\end{subfigure}
	\begin{subfigure}{0.32\textwidth}
		\centering
		\begin{tikzpicture}[baseline,scale=1]

\pgfplotstableread{data/density/density_inf-random-med_acc.csv}\acctable
\pgfplotstableread{data/density/density_inf-random-prct75_acc.csv}\prcttop
\pgfplotstableread{data/density/density_inf-random-prct25_acc.csv}\prctbot

\pgfmathsetmacro{\opacity}{0.2}
\pgfmathsetmacro{\contourop}{0.25}

\begin{axis}[
    xlabel={(c) Edge probability},
    xmin=.1,
    xmax=.7,
    xtick = {.1, .3, ..., .7},
    ylabel={Accuracy},
    ymin = 0,
    ymax = 1.03,
    ytick = {0, .2, ..., 1},
    grid style=densely dashed,
    grid=both,
    legend style={
        at={(.5, 1.02)},
        anchor=south},
    legend columns=3,
    width=175,
    height=165,
    ]

    \addplot [blue!90!black, name path = DCN-bot, opacity=\contourop, forget plot] table [x=xaxis, y=DCN] \prctbot;
    \addplot [blue!90!black, name path = DCN-top, opacity=\contourop, forget plot] table [x=xaxis, y=DCN] \prcttop;
    \addplot[blue!90!black, fill opacity=\opacity, forget plot] fill between[of=DCN-bot and DCN-top];
    \addplot[blue!80!black, mark=o, solid] table [x=xaxis, y=DCN] {\acctable};

    \addplot [red!90!white, name path = GCNN-5-bot, opacity=\contourop, forget plot] table [x=xaxis, y=FB-GCNN-5] \prctbot;
    \addplot [red!90!white, name path = GCNN-5-top, opacity=\contourop, forget plot] table [x=xaxis, y=FB-GCNN-5] \prcttop;
    \addplot[red!90!white, fill opacity=\opacity, forget plot, forget plot] fill between[of=GCNN-5-bot and GCNN-5-top];
    \addplot[red, mark=+, solid] table [x=xaxis, y=FB-GCNN-5] {\acctable};

    \addplot [green!70!black,name path = GraphSAGE-bot, opacity=\contourop, forget plot] table [x=xaxis, y=GraphSAGE-A] \prctbot;
    \addplot [green!70!black,name path = GraphSAGE-top, opacity=\contourop, forget plot] table [x=xaxis, y=GraphSAGE-A] \prcttop;
    \addplot[green!70!black, fill opacity=\opacity, forget plot, forget plot] fill between[of=GraphSAGE-bot and GraphSAGE-top];
    \addplot[green!65!black, mark=x, solid] table [x=xaxis, y=GraphSAGE-A] {\acctable};

    \addplot [blue!60!white, name path = DCN-20-bot, opacity=\contourop, forget plot] table [x=xaxis, y=DCN-20] \prctbot;
    \addplot [blue!60!white, name path = DCN-20-top, opacity=\contourop, forget plot] table [x=xaxis, y=DCN-20] \prcttop;
    \addplot[blue!60!white, fill opacity=\opacity, forget plot] fill between[of=DCN-20-bot and DCN-20-top];
    \addplot[blue!70!white, mark=o, solid] table [x=xaxis, y=DCN-20] {\acctable};

    \addplot [red!60!white, name path = GCNN-2-bot, opacity=\contourop, forget plot] table [x=xaxis, y=FB-GCNN-2] \prctbot;
    \addplot [red!60!white, name path = GCNN-2-top, opacity=\contourop, forget plot] table [x=xaxis, y=FB-GCNN-2] \prcttop;
    \addplot[red!60!white, fill opacity=\opacity, forget plot, forget plot] fill between[of=GCNN-2-bot and GCNN-2-top];
    \addplot[red!70!white, mark=+, dotted] table [x=xaxis, y=FB-GCNN-2] {\acctable};
    
    \addplot [orange!80!black, name path = GIN-bot, opacity=\contourop, forget plot] table [x=xaxis, y=GIN-A] \prctbot;
    \addplot [orange!80!black, name path = GIN-top, opacity=\contourop, forget plot] table [x=xaxis, y=GIN-A] \prcttop;
    \addplot[orange!80!white, fill opacity=\opacity, forget plot, forget plot] fill between[of=GIN-bot and GIN-top];
    \addplot[orange, mark=Mercedes star, solid] table [x=xaxis, y=GIN-A] {\acctable};

    \legend{DCN, FB-GCNN-5, GraphSAGE, DCN-20, FB-GCNN-2, GIN}
\end{axis}
\end{tikzpicture}
	\end{subfigure}
	\caption{(a) NMSE in the network diffusion task as the noise in the observations increases. For the source identification task, we plot accuracy as a function of (b) the proportion of unobserved nodes and (c) the edge probability. We report the median performance across 25 realizations and values between the first and third quartile in the shaded area.}
    \label{fig:exp}
\end{figure*}

%
%

\subsection{Synthetic Data Experiments}\label{ssec:synthetic_data}
{Unless otherwise stated, for the synthetic experiments we sample graphs from an Erd\H{o}s-R\'{e}nyi (ER) random DAG model with $N = 100$ nodes, edge probability $p = 0.2$, and edge weights sampled from a uniform distribution over $[0.2, 1]$.}
We simulate $2000$ input-output graph signal pairs using the model $\bby = \bbH\bbx$. Here, $\bbx$ is a sparse input signal with non-zero values restricted to the first 25 nodes in the DAG, $\bbH$ is a causal graph filter as in (\ref{eq:causal_gf}) constructed from 25 causal GSOs chosen uniformly at random, and $\bby$ represents the diffused output signal.
{The filter taps are independently sampled from a uniform distribution in $[-1, 1]$. After generation, each output signal $\bby$ is normalized to have unit $\ell_2$ norm.}

We focus on two supervised learning problems, involving the above signals supported on ER DAGs; recall the general problem statement in Section~\ref{ssec:prob_statement}.  First, we consider the system identification task of learning a network diffusion process, where the objective is to estimate the output $\bby_{\textrm{test}}$ given a new sparse input $\bbx_{\textrm{test}}$. The input and output signal observations are corrupted with zero-mean additive white Gaussian noise. To have a better control over the signal-to-noise ratio, additive noise is normalized with respect to the signal power. 
We set the normalized power to $0.05$, that is, a $5 \%$ of the power of the signal.
The second task addresses the source identification problem, where the goal is to predict the source node of a partially observed output signal $\bby_{\textrm{test}}$. For this task the input $\bbx$ contains a single non-zero entry, constrained to be within the first 25 nodes in the topologically-ordered $\ccalV$. Since the diffusion $\bby = \bbH\bbx$ does not alter the signal value at the source nodes, we mask the output $\bby$ to exclude direct observations from potential seed nodes. {Results averaged over 25 trials, each corresponding to a different graph realization, are reported in Table~\ref{t:table1}.}\vspace{2pt}

\noindent \textbf{Findings and discussion.} The performance of the (P)DCN architectures is compared to several baselines.
{The least-squares (LS) predictor captures the linear diffusion model}; it computes an LS estimate of $\bbtheta$ from the training data and then predicts test outputs. The results tabulated in Table I show the superior performance of DCN in both tasks. In the \textit{network diffusion task}, we observe that using a subset of 30 or 10 GSOs (DCN-30 and DCN-10) allows {one} to decouple the number of DCN parameters from the graph size, while maintaining competitive performance.
{These subsets of GSOs are selected independently via uniform random sampling from the full set $\ccalV$, and different subsets are used across realizations. Consequently, using fewer GSOs results in higher variability in performance.}
Among these, DCN-30 achieves the second-best results. {Compared to LS, the DCN demonstrates greater resilience to noise and rescaling in the input signals, which we will explore further in a subsequent test case}. Furthermore, when the transposed causal GSOs $\bbS_k^\top$ are used (as in DCN-T), the performance of the model decreases substantially, particularly when only a subset of the GSOs is employed.

In the \textit{source identification task}, the DCN variants struggle to identify the source node, while the transposed DCN-T variants achieve nearly flawless accuracy. This finding reflects the intuitive fact that identifying the source node requires navigating the DAG in the reverse direction, which is better facilitated by the transposed GSOs. This underscores the importance of properly exploiting the directionality of the DAG.  Since the PDCN architecture employs a single-layer MLP with 128 neurons, whereas the DCN utilizes a learnable parameter vector of size 32 in each of its $L=2$ layers, it is expected that the PDCN requires more time for training and inference—as is indeed observed. However, when both architectures are configured with the comparable number of neurons and learnable parameters, their training and inference times match (results not shown here).

Across both tasks, we find that the DAG-agnostic models typically exhibit degraded performance relative to (P)DCN(-T), DAGNN and D-VAE. Interestingly, DCN(-T) exhibits superior performance along with significant savings in training and inference time relative to DAGNN and D-VAE. {This computational advantage stems from the fact that DAGNN and D-VAE rely on more complex parameterizations and sequential processing to account for the DAG structure, whereas DCN encodes directionality directly through causal GSOs. Furthermore, DAG+NodeFormer, which was designed for node classification on citation networks, performs well on the source identification task but struggles on the diffusion learning task, suggesting that its architecture is better aligned with classification objectives than signal regression.}
\vspace{2pt} 

\noindent \textbf{Test Case 1 (Noise Robustness).} For the network diffusion task, Fig.~\ref{fig:exp}(a) examines the impact of increasing noise levels on model performance. 
{The abscissa shows the normalized noise power, while the ordinate reports the NMSE.
In the absence of noise, LS achieves the lowest NMSE among all methods, as it directly implements the linear diffusion model. Perfect recovery is not attained due to the output normalization, which introduces a non-linear transformation. 
As the noise power increases, however, the performance of LS deteriorates rapidly.}
{This is reflected in the steeper slope of the LS curve as noise grows.}
In contrast, the proposed DCN attains an NMSE similar to LS under noiseless conditions but exhibits enhanced robustness in the presence of noise. 
Even though the true underlying generative model is linear, the use of a nonlinear approach like DCN enables enhanced noise tolerance (especially when the sample size is not excessive).\vspace{2pt}

\noindent \textbf{Test Case 2 (Number of Observed Nodes).} 
{For the source identification task, Fig.~\ref{fig:exp}(b) examines the accuracy of DCN as the proportion of nodes that remain unobserved increases. Note that the unobserved nodes are precisely the candidate source nodes.}
In this scenario, higher values along the x-axis correspond to a more challenging setting, as fewer node observations are available and a larger number of unobserved nodes could potentially act as sources. 
Based on the results of Table~\ref{t:table1}, all the methods use the transpose of their respective GSO. The plots show that the DCN maintains high accuracy even when a significant fraction of nodes is unobserved, demonstrating its robustness.
{Notably, the gap between DCN and the remaining methods becomes more pronounced as the number of candidate source nodes grows.}
Among the DCN variants, the performance of DCN-30 and DCN-20 reveals a trade-off between reducing the number of GSOs and maintaining accuracy.
{This effect becomes more pronounced at high levels of unobserved nodes, where reducing the number of GSOs constrains the expressive capacity of the model.}
When $50\%$ or more of the nodes are observed, DCN-30 performs comparably to the full DCN, while DCN-20 outperforms other non-DAG-based methods. Together with Test Case 1, these findings highlight the ability of DCN to balance computational efficiency and predictive accuracy under varying levels of data uncertainty (through noisy or partially-observed signals).\vspace{2pt}

\begin{figure}
    \centering
    \begin{tikzpicture}[baseline,scale=1]

\pgfplotstableread[col sep=semicolon]{data/mult_src_id/exp_mult_src_sweep-med_rec.csv}\medtable
\pgfplotstableread[col sep=semicolon]{data/mult_src_id/exp_mult_src_sweep-prct25_rec.csv}\prctbot
\pgfplotstableread[col sep=semicolon]{data/mult_src_id/exp_mult_src_sweep-prct75_rec.csv}\prcttop

\pgfmathsetmacro{\opacity}{0.30}
\pgfmathsetmacro{\contourop}{0.25}

\begin{axis}[
    xlabel={Number of source nodes},
    xmin=1, xmax=5,
    xtick={1,2,3,4,5},
    ylabel={Median Recall},
    ymin=0, ymax=1.03,
    ytick={0,.2,...,1},
    grid=both,
    grid style=densely dashed,
    legend style={
        at={(1.02,0.5)},
        anchor=west,
    },
    legend columns=1,
    width=200,
    height=135,
]

\addplot [blue!90!black, name path=DCNT-bot, opacity=\contourop, forget plot]
    table [x=xaxis, y=DCN-T] {\prctbot};
\addplot [blue!90!black, name path=DCNT-top, opacity=\contourop, forget plot]
    table [x=xaxis, y=DCN-T] {\prcttop};
\addplot [blue!90!black, fill opacity=\opacity, forget plot]
    fill between[of=DCNT-bot and DCNT-top];
\addplot [blue!80!black, mark=o, solid]
    table [x=xaxis, y=DCN-T] {\medtable};

\addplot [blue!90!white, name path=DCN30T-bot, opacity=\contourop, forget plot]
    table [x=xaxis, y=DCN-30-T] {\prctbot};
\addplot [blue!90!white, name path=DCN30T-top, opacity=\contourop, forget plot]
    table [x=xaxis, y=DCN-30-T] {\prcttop};
\addplot [blue!90!white, fill opacity=\opacity, forget plot]
    fill between[of=DCN30T-bot and DCN30T-top];
\addplot [blue, mark=o, densely dotted]
    table [x=xaxis, y=DCN-30-T] {\medtable};

\addplot [red!80!black, name path=FBGCNN4-bot, opacity=\contourop, forget plot]
    table [x=xaxis, y=FB-GCNN-4] {\prctbot};
\addplot [red!80!black, name path=FBGCNN4-top, opacity=\contourop, forget plot]
    table [x=xaxis, y=FB-GCNN-4] {\prcttop};
\addplot [red!80!white, fill opacity=\opacity, forget plot]
    fill between[of=FBGCNN4-bot and FBGCNN4-top];
\addplot [red, mark=+, solid]
    table [x=xaxis, y=FB-GCNN-4] {\medtable};

\addplot [green!70!black, name path=SAGE-bot, opacity=\contourop, forget plot]
    table [x=xaxis, y=GraphSAGE-A] {\prctbot};
\addplot [green!70!black, name path=SAGE-top, opacity=\contourop, forget plot]
    table [x=xaxis, y=GraphSAGE-A] {\prcttop};
\addplot [green!70!black, fill opacity=\opacity, forget plot]
    fill between[of=SAGE-bot and SAGE-top];
\addplot [green!65!black, mark=x, solid]
    table [x=xaxis, y=GraphSAGE-A] {\medtable};


\legend{DCN-T, DCN-30-T, FB-GCNN-4, GraphSAGE-A}

\end{axis}
\end{tikzpicture}
    \caption{{Performance in the source identification task as the number of source nodes increases. We report the median recall, measured as the ratio of correctly identified source nodes, and the values between the first and third quartiles.}}
    \label{fig:mult_src_exp}
\end{figure}

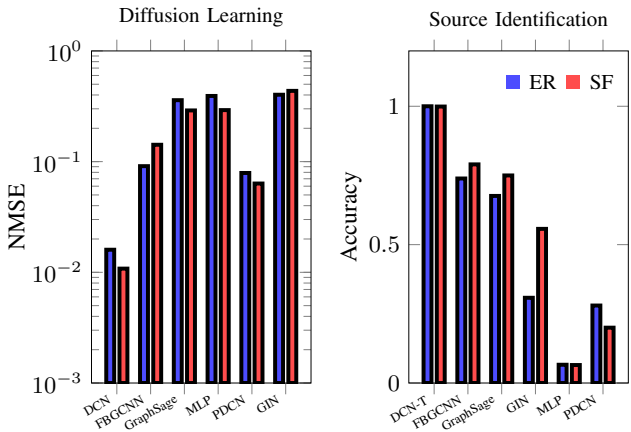
\begin{figure}[t]
\centering

\begin{tikzpicture}

\begin{groupplot}[
    group style={
        group size=2 by 1,
        horizontal sep=1.3cm,
    },
    width=4.5cm,
    height=6cm,
    ybar,
    ymin=0,
    xticklabel style={font=\tiny, rotate=30, anchor=east},
    enlarge x limits=0.15,
    every node near coord/.append style={font=\tiny},
]

\nextgroupplot[
    title={\footnotesize Diffusion Learning},
    ymin=0.001, ymax=1,
    ymode=log,
    ylabel={NMSE},
    line width=0.02pt,
    bar width=3pt, 
    log origin=infty,
    symbolic x coords={DCN, FBGCNN, GraphSage, MLP, PDCN, GIN},
    xtick=data
]
\addplot[fill=blue!70] coordinates {
    (DCN, 0.016) (FBGCNN, 0.091) (GraphSage, 0.359) (MLP, 0.392) (PDCN, 0.079) (GIN, 0.402)
};
\addplot[fill=red!70] coordinates {
    (DCN, 0.0108) (FBGCNN, 0.142) (GraphSage, 0.290) (MLP, 0.292) (PDCN, 0.0634) (GIN, 0.4357)
};

\nextgroupplot[
    title={\footnotesize Source Identification},
    ymin=0, ymax=1.2,
    ylabel={Accuracy},
    line width=0.02pt,
    bar width=3pt, 
    symbolic x coords={DCN-T, FBGCNN, GraphSage, GIN, MLP, PDCN},
    xtick=data
]
\addplot[fill=blue!70] coordinates {
    (DCN-T, 1) (FBGCNN, 0.739) (GraphSage, 0.676) (GIN, 0.308) (MLP, 0.066) (PDCN, 0.28)
};
\addplot[fill=red!70] coordinates {
    (DCN-T, 0.999) (FBGCNN, 0.790) (GraphSage, 0.750) (GIN, 0.557) (MLP, 0.065) (PDCN, 0.2)
};

\end{groupplot}

\node[draw=none, fill=blue!70, minimum size=5pt, inner sep=1pt] at (5.6,4) {};
\node[anchor=west, font=\footnotesize] at (5.7,4) {ER};

\node[draw=none, fill=red!70, minimum size=5pt, inner sep=1pt] at (6.4,4) {};
\node[anchor=west, font=\footnotesize] at (6.5,4) {SF};

\end{tikzpicture}

\vspace{0.2cm}

\caption{Comparison of Erd\H{o}s–R\'enyi (ER) and scale-free (SF) graphs in diffusion learning (left) and source identification (right). Performance is fairly invariant across graph types.}
\label{fig:clustered_bar_plot}
\end{figure}


\noindent \textbf{Test Case 3 (Edge Density).} We also assess the impact of the graph's edge density on the source identification problem. To this end, for various models including DCN we evaluate accuracy as the ER edge probability $p$ increases. 
{Here, larger values of $p$ correspond to denser graphs with a higher number of overlapping directed paths.}
Figure~\ref{fig:exp}(c) shows that as the graph becomes denser, identifying the source node becomes more challenging. This is because denser graphs tend to generate more homogeneous signals, with differences primarily arising from the distance between source nodes.
{As connectivity intensifies, signals propagated from different sources become less distinguishable.}
Notice how both the DCN and its variant DCN-20 consistently outperform FB-GCNN-$R$, which uses filters of order $R=2$ and $5$.
{The gap is particularly evident in the intermediate-density regime, where the FB-GCNN variants exhibit a steeper performance decay than DCN.}
The results corroborate the benefits of adopting convolutional filters on DAGs over legacy graph filters proposed in the GSP literature. Even as the graph becomes denser, the DCN performance degrades gracefully.\vspace{2pt} 

{
\noindent \textbf{Test Case 4 (Number of Source Nodes).} 
Next, we consider a source identification setting where multiple nodes simultaneously act as signal sources. The number of source nodes is assumed to be known.
For a subset of models, Figure~\ref{fig:mult_src_exp} reports the performance as the number of active source nodes increases. The vertical axis shows the recall, defined as the ratio of correctly identified source nodes to the total number of true sources.
As expected, the difficulty of the problem increases with the number of sources.
We observe that DCN and DCN-30 achieve almost perfect recall when there is a single source node, consistent with previous experiments, and still identify roughly half of the sources as the complexity increases.
This performance degradation reflects the inherently challenging nature of the task.
Moreover, the proposed architectures consistently outperform the baselines, which struggle to identify the source nodes, highlighting the benefits of our DAG-based GNNs even in demanding settings. \vspace{2pt}
}

\noindent \textbf{Test Case {5} (Graph Type).} Figure~\ref{fig:clustered_bar_plot} compares the performance of six models— DCN, FB-GCNN, GraphSAGE, MLP, PDCN, and GIN—on two tasks: diffusion learning (left) and source identification (right), where the problem instances were generated for ER and scale-free (SF)~\cite{barabasi1999emergence} DAGs. In the diffusion learning task, the results are shown on a logarithmic scale to accommodate the wide dynamic range of NMSE values. Among the models, DCN consistently achieves the lowest NMSE, on both the SF and ER graphs. In addition, PDCN performs competitively, with fewer parameters. Moving on to source identification, the results in Fig.~\ref{fig:clustered_bar_plot}(right) indicate that DCN-T is able to identify the true source with high accuracy, outperforming the other models in both SF and ER networks. This superior performance highlights the robustness and generalization of DCN across various random DAG ensembles. \vspace{2pt} 

\begin{table*}[t]
\begingroup
\centering
{
\captionsetup{font=small}
\resizebox{\textwidth}{!}{
\tiny

\begin{tabular}{l c c c c c c c c}
\toprule
\textbf{Method} & \multicolumn{2}{c}{$N=100$} & \multicolumn{2}{c}{$N=200$} & \multicolumn{2}{c}{$N=500$} & \multicolumn{2}{c}{$N=1000$} \\
\cmidrule(lr){2-3}\cmidrule(lr){4-5}\cmidrule(lr){6-7}\cmidrule(lr){8-9}
 & \textbf{NMSE} & \textbf{Time} & \textbf{NMSE} & \textbf{Time} & \textbf{NMSE} & \textbf{Time} & \textbf{NMSE} & \textbf{Time} \\
\midrule
DCN             & $\mathbf{0.0065 \pm 0.0043}$ & 26.0 & $\mathbf{0.0083 \pm 0.0046}$ & 140.8 & $\mathbf{0.0058 \pm 0.0056}$ & 4732.7 & $\mathbf{0.0073 \pm 0.0038}$ & 25604.8 \\
DCN-50         & $0.0156 \pm 0.0110$ & 5.8 & $0.0151 \pm 0.0072$ & 9.5 & $0.0114 \pm 0.0138$ & 45.6 & $0.0092 \pm 0.0092$ & 855.5 \\
PDCN            & $0.0778 \pm 0.0180$          & 141.5 & $0.0988 \pm 0.0181$          & 370.2 & $0.1621 \pm 0.0984$ & 4241.9 & $1.6522 \pm 0.2088$ & 14357.6 \\
I-PDCN & $0.0263 \pm 0.0115$ & 166.1 & $0.0948 \pm 0.0416 $ & $326.1$ & $0.1202 \pm 0.1103$ & $1352.7$ & $0.2636 \pm 0.1730$ & $3251.0$ \\
\midrule
DAGNN           & $0.1014 \pm 0.0201$ & 1805.5 & $0.1268 \pm 0.0205$ & 5860.2 & DNF & DNF & DNF & DNF \\
D-VAE           & $0.1017 \pm 0.0201$ & 10155.3 & $0.1274 \pm 0.0210$ & 13230.2 & DNF & DNF & DNF & DNF \\
DAG+NodeFormer       & $0.5810 \pm 0.0165$ & 744.1 & $0.6635 \pm 0.0187$ & 1043.2 & DNF & DNF & DNF & DNF \\
DAG+SAT       & $0.2727 \pm 0.1509$ & 2283.0 & $0.3309 \pm 0.0928$ & 2842.6 & DNF & DNF & DNF & DNF \\
\midrule
Linear          & $0.0211 \pm 0.0134$          & 0.4 & $0.0188 \pm 0.0092$          & $1.4$ & $0.0162 \pm 0.0181$ & 34.4 & $0.0174 \pm 0.0166$ & 228.5 \\
Scarselli-GNN   & $0.0330 \pm 0.0065$ & 172.5 & $0.0698 \pm 0.0048$ & $179.4$ & $0.2421 \pm 0.1339$ & 1573.0 & $0.6870 \pm 0.7065$ & 6101.6 \\
GCN             & $0.2008 \pm 0.0254$ & 6.3 & $0.2982 \pm 0.0226$ & $6.4$ & $0.3346 \pm 0.1129$ & 6.5 & $0.4202 \pm 0.0794$ & 7.8 \\
GraphSAGE       & $0.5776 \pm 0.0165$ & 13.3 & $0.6609 \pm 0.0180$ & $13.4$ & $0.7177 \pm 0.1613$ & 11.7 & $0.6743 \pm 0.2649$ & 12.1 \\
GIN             & $0.6107 \pm 0.0286$ & 12.4 & $0.7354 \pm 0.0183$ & $12.3$ & $0.8408 \pm 0.0727$ & 12.9 & $0.9172 \pm 0.0927$ & 23.3 \\
GAT             & $0.6445 \pm 0.0191$ & 18.7 & $0.7560 \pm 0.0074$ & $16.0$ & $0.8337 \pm 0.0729$ & 16.9 & $0.8821 \pm 0.0559$ & 40.9 \\
GGNN            & $0.5309 \pm 0.0191$ & 405.0 & $0.7077 \pm 0.0102$ & $471.5$ & $0.7799 \pm 0.1181$ & 497.5 & $0.8273 \pm 0.1128$ & 412.5 \\
\midrule
MLP             & $0.5803 \pm 0.0161$ & 3.1 & $0.6621 \pm 0.0161$ & $3.2$ & $0.7162 \pm 0.1630$ & 4.0 & $0.6700 \pm 0.2712$ & 3.6 \\
GRU             & $0.3041 \pm 0.2216$ & 13.3 & $0.3035 \pm 0.1749$ & 14.6 & $0.7733 \pm 0.1627$ & 14.0 & $0.7220 \pm 0.2135$ & 24.4 \\
LSTM            & $0.2961 \pm 0.2151$ & 13.3 & $0.2559 \pm 0.1495$ & 15.0 & $0.7487 \pm 0.1630$ & 15.8 & $0.7348 \pm 0.2090$ & 24.9 \\
\bottomrule
\end{tabular}
}
\caption{{NMSE and training time for the diffusion learning task as the graph size $N$ increases, averaged over 10 independent trials.}}
\label{tab:diffusion_scaling}
}
\endgroup
\end{table*}

{\noindent \textbf{Test Case 6 (Scalability).} 
The final synthetic experiment evaluates the scalability of the proposed methods.
Table~\ref{tab:diffusion_scaling} reports the NMSE in the diffusion learning task as the number of nodes increases from $N{=}100$ to $N{=}1000$. Entries marked as DNF (Did Not Finish) indicate models that exceeded the available computation time.
DCN, which uses all GSOs, achieves the lowest errors for smaller graphs but encounters memory limitations as the graph size increases, leading to longer training times for $N=1000$. When restricted to a subset of 50 GSOs, DCN-50 maintains low NMSE while achieving reasonable training times across all graph sizes. PDCN employs a single shared MLP across all branches, whereas I-PDCN assigns an independent MLP to each branch and uses only 50 GSOs. This additional expressivity proves beneficial, as I-PDCN consistently outperforms PDCN on larger graphs.
As in previous tests, DAG-aware baselines do not scale well due to prohibitive training times. DAG+NodeFormer yields high errors, which is expected since it was designed for node classification rather than signal regression, while DAG+SAT~\cite{luo2023transformers} performs comparatively better thanks to its structure-aware attention mechanism. GNN models that ignore the DAG structure also exhibit higher errors, particularly for larger graphs.
Overall, these results indicate that the proposed architectures scale effectively, offering a favorable trade-off between computational complexity and predictive accuracy.}

\subsection{Imputing Missing Gene-Expression Levels}\label{ssec:genes_dataset}

We use a real-world gene expression dataset from \textit{Arabidopsis Thaliana} \cite{opgenrhein2007correlation}, collected to study the effect of the diurnal cycle on starch metabolism. The data consist of time series measurements for 22,814 genes across 11 time points in two biological replicates, available from the NASCArrays repository. After preprocessing (log transformation, filtering low-expression and incomplete entries), 800 genes showing diurnal periodicity were selected.
To infer causal structure, partial correlation testing identified 6,102 significant edges among 669 genes. Directionality was determined for a subset using a likelihood-based test, yielding 1,216 significant directions. We focus on a 107-node, 150-edge subgraph containing the strongest directed associations. This DAG displays a hub-like topology, with key regulatory genes such as transcription factors showing mostly outgoing edges, consistent with known biological roles. To ensure the stability of the inferred network, a robustness analysis was carried out, confirming that the edges were not artifacts of noise or random variability~\cite{opgenrhein2007correlation}.

\begin{table}[t]
    \captionsetup{font=small}
    \centering
    \resizebox{\columnwidth}{!}{
\begin{tabular}{l|c c|c c|c} 
    \multicolumn{1}{c|}{\textbf{ Model}} 
    & \multicolumn{2}{c|}{\textbf{ Masked Nodes: 70}} 
    & \multicolumn{2}{c|}{\textbf{ Masked Nodes: 80}} 
    & \textbf{ Parameter} \\ 
    & \textbf{ NMSE} & \textbf{ Time (s)} 
    & \textbf{ NMSE} & \textbf{ Time (s)} 
    & \textbf{ Count} \\
    \hline
    DCN & $0.025 \pm 0.002$ & 3.3 & $0.027 \pm 0.003$ & 3.4 & 6881 \\
    PDCN & $0.052 \pm 0.008$ & 1.4 & $0.053 \pm 0.008$ & 1.6 & 385 \\
    \hline
    D-VAE & $\mathbf{0.010 \pm 0.003}$ & 72.3 & $\mathbf{0.011 \pm 0.002}$ & 84.1 & 450669 \\
    DAGNN & $0.014 \pm 0.003$ & 23.7 & $0.014 \pm 0.007$ & 23.5 & 917997 \\
    {DAG+NodeFormer} & {$0.054 \pm 0.008$} & {30.3} & {$0.055 \pm 0.008$} & {25.1} & {158657} \\
    \hline
    {LS} & $0.103 \pm 0.003$ & $\leq 0.0$ & $0.217 \pm 0.003$ & $\leq 0.0$ & 107 \\
    GNN-A & $0.081 \pm 0.007$ & 3.1 & $0.077 \pm 0.007$ & 3.1 & 97 \\
    FB-GCNN-2 & $0.033 \pm 0.005$ & 3.4 & $0.036 \pm 0.005$ & 3.5 & 161 \\
    GraphSAGE-A & $0.031 \pm 0.005$ & 6.2 & $0.034 \pm 0.005$ & 6.3 & 161 \\
    GIN & $0.039 \pm 0.007$ & 5.6 & $0.040 \pm 0.003$ & 5.9 & 2209 \\
    GIN-4 & $0.031 \pm 0.004$ & 7.3 & $0.032 \pm 0.002$ & 7.3 & 6433 \\
    MLP & $0.036 \pm 0.006$ & 2.2 & $0.039 \pm 0.006$ & 2.2 & 97 \\
    GAT & $0.045 \pm 0.008$ & 12.5 & $0.045 \pm 0.008$ & 12.6 & 163 \\
    {GRU} & {$0.035 \pm 0.008$} & {3.2} & {$0.042 \pm 0.009$} & {3.2} & {9729} \\
    {LSTM} & {$0.038 \pm 0.009$} & {3.4} & {$0.040 \pm 0.009$} & {3.4} & {12961} \\
    {GGNN} & {$0.017 \pm 0.001$} & {25.3} & {$0.021 \pm 0.002$} & {30.0} & {7489} \\
    {Scarselli-GNN} & {$0.029 \pm 0.004$} & {11.2} & {$0.038 \pm 0.005$} & {7.0} & {969} \\
    \hline
\end{tabular}
    }
    \caption{Performance comparison of various models on the Arabidopsis Thaliana gene network for node feature prediction. Results are reported for two masking scenarios, with 70 and 80 nodes masked during training. For each model, the NMSE and training plus testing time (in seconds) over 25 trials are reported. The rightmost column indicates the number of trainable parameters. The lowest NMSE values for each target are highlighted in bold.}
    \label{tab:model_performance}
\end{table}

Gene expression measurements recorded at 11 time points across two biological replicates are treated as 22 graph signals defined over the DAG, where $N=107$. We consider the imputation task of interpolating missing gene-expression level observations in a subset of nodes. Given partially observed graph signals $\bbx$ where measurements on some of the nodes are unavailable (or masked), the goal is to predict those missing expression levels.
{Hence, the input-output pairs consist of the masked signals as inputs and the corresponding full signals as outputs.}
During training, the graph signal $\bby$ includes the target missing observations as a supervision signal. The models are trained and validated on separate subsets of the data, and their performance is assessed on the test set using the signal reconstruction NMSE. The results, summarized in Table~\ref{tab:model_performance}, demonstrate the strong performance of the proposed DCN model, which comes close to the best performing DAGNN and D-VAE and outperforms all other methods. 
{Interestingly, GGNN attains competitive performance despite not being tailored to DAGs.}
The PDCN variant also yields competitive results, closely following the leading baselines.
{To evaluate robustness, the experiment is repeated with 70 and 80 masked nodes, consistently showing that DCN maintains satisfactory predictive accuracy despite increasing amounts of missing data.}
Additionally, the number of parameters used by each model is reported to assess computational efficiency. Notably, DAGNN and D-VAE incur markedly higher memory and runtime costs due to their architectural complexity. In contrast, PDCN achieves favorable performance with far fewer parameters—scaling independently of the number of nodes. While the additional time required to train DAGNN and D-VAE is not prohibitive in this test case where the dataset is small, for larger problems the efficiency gains of (P)DCN can be significant (see also {the scalability tests in Table~\ref{tab:diffusion_scaling}}).

\begin{table*}[t]
    \centering
    \captionsetup{font=small}
    \resizebox{\textwidth}{!}{ 
\begin{tabular}{l|c c | c c | c c | c c| c} 
    & \multicolumn{2}{c|}{\textbf{Dissolved Reactive Silicon}}
    & \multicolumn{2}{c|}{\textbf{Soluble Sulphate}}
    & \multicolumn{2}{c|}{\textbf{Dissolved Nitrate}}
    & \multicolumn{2}{c|}{\textbf{Dissolved Chloride}} & Parameter\\ 
    & NMSE & Time (s) & NMSE & Time (s) & NMSE & Time (s) & NMSE & Time (s) & Count \\ 
    \hline
    DCN & $\mathbf{0.006 \pm 0.010}$ & 2.26 & $\mathbf{0.003 \pm 0.003}$ & 2.20 & $\mathbf{0.004 \pm 0.003}$ & 2.27 & $\mathbf{0.020 \pm 0.100}$ & 2.26 & 1313 \\
    DCN-15 & $0.049 \pm 0.020$ & 2.28 & $0.031 \pm 0.007$ & 2.19 & $0.073 \pm 0.020$ & 2.27 & $0.076 \pm 0.096$ & 2.26 & 993\\
    PDCN & $0.037 \pm 0.071$ & 1.77 & $0.109 \pm 0.015$ & 1.74 & $0.120 \pm 0.052$ & 1.77 & $0.129 \pm 0.095$ & 1.76 & 385 \\
    DCN-T & $0.054 \pm 0.064$ & 2.10 & $0.116 \pm 0.028$ & 2.23 & $0.069 \pm 0.022$ & 2.29 & $0.117 \pm 0.097$ & 2.29 & 1313 \\
    \hline
    D-VAE & $0.012 \pm 0.010$ & 183.42 & $0.033 \pm 0.003$ & 177.61 & $0.086 \pm 0.043$ & 182.79 & $0.040 \pm 0.100$ & 183.44 & 460694 \\
    DAGNN & $0.025 \pm 0.028$ & 144.89 & $0.044 \pm 0.010$ & 112.80 & $0.086 \pm 0.021$ & 84.43 & $0.049 \pm 0.098$ & 144.85 & 888454 \\
    {DAG+NodeFormer}      
    & {$0.059 \pm 0.042$} & {$434.3$} 
    & $0.174 \pm 0.097$ & {$346.62$} 
    & $0.105 \pm 0.039$ & {$447.42$} 
    & $0.104 \pm 0.017$ & {$398.53$} & {40417}\\
    \hline
    LS & $0.009 \pm 0.015$ & 0.01 & $0.004 \pm 0.003$ & 0.01 & $0.006 \pm 0.005$ & 0.01 & $0.022 \pm 0.099$ & 0.01 & 20 \\
    LS-15 & $0.096 \pm 0.032$ & 0.01 & $0.329 \pm 0.028$ & 0.01 & $0.006 \pm 0.005$ & 0.01 & $0.022 \pm 0.099$ & 0.01 & 15 \\
    GraphSAGE-A & $0.029 \pm 0.036$ & 3.90 & $0.022 \pm 0.012$ & 3.81 & $0.022 \pm 0.018$ & 3.89 & $0.040 \pm 0.101$ & 3.89 & 161 \\
    GIN-A & $0.057 \pm 0.056$ & 3.87 & $0.174 \pm 0.028$ & 3.73 & $0.204 \pm 0.031$ & 3.83 & $0.192 \pm 0.074$ & 3.83 & 33409 \\
    GIN-A-4 & $0.036 \pm 0.049$ & 4.60 & $0.171 \pm 0.028$ & 4.80 & $0.194 \pm 0.033$ & 4.98 & $0.186 \pm 0.075$ & 4.99 & 99457 \\
    MLP & $0.073 \pm 0.103$ & 1.63 & $0.171 \pm 0.028$ & 1.58 & $0.126 \pm 0.033$ & 1.62 & $0.159 \pm 0.099$ & 1.62 & 97 \\
    GNN-A & $0.537 \pm 0.077$ & 2.01 & $0.628 \pm 0.040$ & 1.96 & $0.670 \pm 0.059$ & 2.00 & $0.657 \pm 0.058$ & 2.02 & 97 \\
    FB-GCNN-2 & $0.046 \pm 0.027$ & 2.27 & $0.010 \pm 0.003$ & 2.21 & $0.021 \pm 0.007$ & 2.24 & $0.032 \pm 0.101$ & 2.25 & 161 \\
    GAT & $0.042 \pm 0.034$ & 7.88 & $0.084 \pm 0.018$ & 7.65 & $0.059 \pm 0.023$ & 7.82 & $0.069 \pm 0.092$ & 7.85 & 163 \\
    {GRU}   
    & {$0.186 \pm 0.196$} & {1.23}
    & {$0.628 \pm 0.083$} & {1.12}
    & {$0.550 \pm 0.075$} & {1.26}
    & {$0.686 \pm 0.029$} & {1.16} & {3393} \\
    {LSTM}     
    & {$0.221 \pm 0.249$} & {1.24} & 
    {$0.622 \pm 0.085$} & {1.15} &
    {$0.541 \pm 0.076$} & {1.18} &
    {$0.674 \pm 0.030$} & {1.17} & {4513}\\
    {GGNN}   
    & {$0.018 \pm 0.020$} & {53.85}
    & {$0.180 \pm 0.097$} & {53.86}
    & {$0.137 \pm 0.023$} & {54.02}
    & {$0.209 \pm 0.029$} & {53.91} & {7489}\\

    {Scarselli-GNN}   
    & {$0.042 \pm 0.032$} & {5.01} 
    & {$0.060 \pm 0.101$} & {3.75} 
    & {$0.030 \pm 0.015$} & {3.56}
    & {$0.060 \pm 0.024$} & {3.80} & {969}\\

    \hline
\end{tabular}
}
    \caption{Performance comparison of various models on the River Thames network for node feature prediction. Results are reported for four different chemical indicators. For each model, the mean NMSE ± standard deviation and mean training plus testing time (in seconds) over 25 trials are reported. The rightmost column indicates the number of trainable parameters. The lowest NMSE values for each target are highlighted in bold.}
    \label{tab:performance}
\end{table*}

%
%

\subsection{Hydrological Data Forecasting}\label{ssec:thames_dataset}
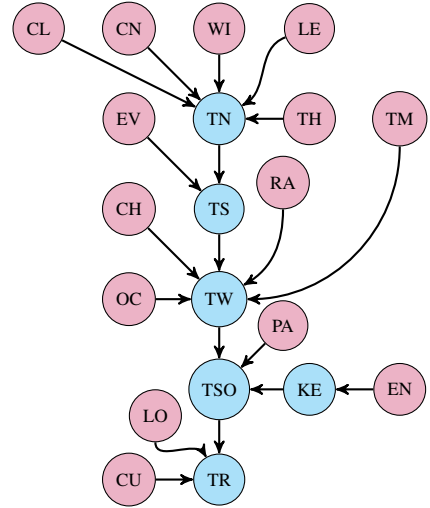
\begin{figure}[t]
    \centering
    \begin{tikzpicture}[
        scale=0.5, 
        ->, >=stealth',
        node distance=1.2cm and 1.5cm, 
        every node/.style={font=\scriptsize}, 
        main/.style={circle, draw, fill=cyan!30, minimum size=0.6cm, text centered},
        source/.style={circle, draw, fill=purple!30, minimum size=0.6cm, text centered}
    ]

    \node[source] (WI) {WI};
    \node[source] (CN) [left of=WI] {CN};
    \node[source] (LE) [right of=WI] {LE};
    \node[main] (TN) [below of=WI] {TN};
    \node[source] (TH) [right of=TN] {TH};
    \node[source] (CL) [left of=CN] {CL};
    \node[source] (RA) [below right of=TN] {RA};
    \node[source] (TM) [right of=TH] {TM};
    \node[source] (EV) [left of=TN] {EV};
    \node[main] (TS) [below of=TN] {TS};
    \node[source] (CH) [left of=TS] {CH};
    \node[main] (TW) [below of=TS] {TW};
    \node[source] (OC) [left of=TW] {OC};
    \node[main] (TSO) [below of=TW] {TSO};
    \node[source] (PA) [above right of=TSO] {PA};
    \node[main] (KE) [right of=TSO] {KE};
    \node[source] (EN) [right of=KE] {EN};
    \node[main] (TR) [below of=TSO] {TR};
    \node[source] (CU) [left of=TR] {CU};
    \node[source] (LO) [above left of=TR] {LO};

    \draw[thick] (WI) -- (TN);
    \draw[thick] (CN) -- (TN);
    \draw[thick] (LE) to[out=200, in=30] (TN);
    \draw[thick] (TH) -- (TN);
    \draw[thick] (CL) -- (TN);
    \draw[thick] (TN) -- (TS);
    \draw[thick] (EV) -- (TS);
    \draw[thick] (TM) to[out=270, in=0] (TW); 
    \draw[thick] (RA) to[out=270, in=30] (TW);
    \draw[thick] (OC) -- (TW);
    \draw[thick] (CH) -- (TW);
    \draw[thick] (TS) -- (TW);
    \draw[thick] (TW) -- (TSO);
    \draw[thick] (TSO) -- (TR);
    \draw[thick] (PA) -- (TSO);
    \draw[thick] (KE) -- (TSO);
    \draw[thick] (EN) -- (KE);
    \draw[thick] (CU) -- (TR);
    \draw[thick] (LO) to[out=270, in=120] (TR); 

    \end{tikzpicture}
    \caption{A DAG representing the flow of the River Thames and the locations of measurement sites along its course. Purple nodes denote source sites, whereas cyan nodes indicate intermediary or sink sites. For the nodal acronym definitions; see~\cite{bowes2020weekly}.}
    \label{fig:dag_thames}
\end{figure}

The River Thames dataset~\cite{bowes2020weekly} is a hydrological and chemical dataset collected weekly from 2009 to 2017 across 20 monitoring sites along the river. It includes measurements such as lab pH, gran alkalinity, suspended solids, reactive soluble phosphorus, and concentrations of dissolved substances such as chloride and metals, along with mean daily flow rates. Due to substantial missing values in the data for 2009 and 2017, tests are conducted using the more complete data from 2010 to 2016. This subset provides a reliable basis for studying signal propagation and testing predictive models in hydrological systems.
For the experiments in this section we consider a DAG where nodes correspond to $N=20$ monitoring sites along the river, and directed edges capture the upstream-to-downstream flow relationships; see Fig.~\ref{fig:dag_thames}. Graph signals studied consist of 343 realizations of chemical concentration measurements recorded at the sites, including measures such as dissolved reactive silicon, soluble reactive sulphate, dissolved chloride, and dissolved nitrate.

We consider a graph signal imputation task as in Section~\ref{ssec:genes_dataset}, where certain nodes are masked to simulate real-world scenarios involving missing data. The masked nodes include intermediary or sink nodes, colored cyan in Fig.~\ref{fig:dag_thames}. The purple source nodes in Fig.~\ref{fig:dag_thames} are fully observed to provide input signals. The interpolation goal is to predict the concentrations at the masked nodes based on the observed signals at the source nodes. The results presented in Table~\ref{tab:performance} illustrate the superior performance of DCN in all four considered chemical measures. In particular, DCN achieves the lowest NMSE values among all methods, corroborating its ability to exploit the DAG structure of the River Thames network. Among the DCN variants, DCN-15 represents a good balance between computational efficiency and predictive accuracy. As the number of GSOs decreases, model performance deteriorates; however, this reduction significantly lowers the number of parameters. In terms of parameter efficiency, training time, and predictive accuracy, PDCN performs competitively, surpassing several baseline methods. Once more, notice DCN's several orders-of-magnitude savings in the amount of parameters as well as training and inference times relative to DAGNN and D-VAE -- without sacrificing predictive performance.

%
%

\section{Conclusions, Limitations, and Future Work}~\label{sec:conclusion}
This work introduced DCN, a novel GNN architecture specifically designed for convolutional learning from signals supported on DAGs. Unlike conventional GNNs, DCN layers are built from causal graph filters that inject strong inductive biases stemming from the partially-ordered nodes in a DAG. Unlike prior message-passing architectures for DAG representation learning, our GSP perspective is to exploit formal convolutions for DAG signals. These sparse operators admit graph-spectral domain interpretations (useful for analyses) and computationally-efficient implementation in the vertex domain. We further proposed PDCN, a parallel and parameter-efficient variant that retains strong predictive performance.
We established that the proposed architectures are permutation equivariant and that PDCN is capable of mapping non-isomorphic DAGs, as recognized by the WL test, to distinct representations. Extensive experiments on synthetic and real-world datasets, including prediction of gene-expression levels and hydrological forecasting, confirmed the robustness, scalability, and effectiveness of the proposed architectures. DCN consistently outperformed state-of-the-art baselines in terms of accuracy, noise resilience, and computational efficiency.

Despite these strengths, our approach is not exempt from limitations. DCN's reliance on a potentially large number of causal GSOs (and accordingly the number of learnable weights that grows with the number of nodes) may challenge scalable learning in moderate- to large-sized DAGs. Although PDCN addresses this issue in part, future work will explore adaptive or learned selection of informative GSOs to further reduce complexity. Additionally, we aim to investigate how DCN can be extended to dynamic DAGs or probabilistic causal graphs; integration within Bayesian neural network frameworks is also of interest for adding an uncertainty quantification substrate to our models. Conducting stability analyses in the presence of graph perturbations is another direction in our future research agenda, noting that studies involving directed graphs are notoriously challenging. Overall, this work contributes a principled and practical foundation for deep learning on DAGs, with broad implications across diverse domains where directional (and potentially causal) relationships are of essence.

%
%

\appendices

%
%

\section{Proof of Proposition~\ref{prop:uniqueness}}\label{a:uniqueness}
{Consider two non-isomorphic DAGs, $\mathcal{D}$ and $\mathcal{D}'$.
Since every DAG admits a topological ordering, we may represent both graphs under such orderings so that their adjacency matrices $\bbA$ and $\bbA'$ are strictly lower triangular.
Because the DAGs are non-isomorphic, no relabeling can make their adjacency matrices coincide; hence, $\bbA \neq \bbA'$.}
The associated causal GSO matrices $\bbS_k$ and $\bbS_k'$ are given by
\begin{align} \label{eq:diff_CGSO}
    \bbS_k = \bbW \bbD_k \bbW^{-1}, \quad \bbS_k' = \bbW' \bbD_k' (\bbW')^{-1},
\end{align}
where $\bbW = (\bbI - \bbA)^{-1}$ and $\bbW' = (\bbI - \bbA')^{-1}$ are the respective transitive closures.
{Note that the operation $\bbW=(\bbI-\bbA)^{-1}$ is injective, since $\bbA=\bbI-\bbW^{-1}$.
Hence, $\bbA \neq \bbA'$ implies $\bbW \neq \bbW'$.}

Next, suppose there exist some matrices $\bbD_k$ and $\bbD_k'$ such that $\bbS_k = \bbS_k'$ even when $\bbW \neq \bbW'$.
In such a case, using \eqref{eq:diff_CGSO} results in the equality
\begin{equation} \label{eq:equality}
    \bbD_k = \bbW^{-1}\bbW' \bbD_k' (\bbW')^{-1} \bbW = \bbS \bbD_k' \bbS^{-1},
\end{equation}
where we define $\bbS := \bbW^{-1} \bbW'$.
Expanding $\bbS$, we have
\begin{equation} \label{eq:S}
    \bbS = (\bbI - \bbA)(\bbI - \bbA')^{-1} = (\bbI - \bbA)\sum_{\ell=0}^{\infty} (\bbA')^{\ell} = \bbI - \bbA + \bbJ, 
\end{equation}
with $\bbJ := (\bbI -\bbA) \sum_{\ell=1}^{\infty} (\bbA')^{\ell}$.
Since $\bbA$ is strictly lower triangular (as it corresponds to a DAG), it follows that $\bbJ$ is also strictly lower triangular.
Similarly, we obtain
\begin{equation} \label{eq:S_inv}
    \bbS^{-1} = (\bbI - \bbA)^{-1} (\bbI - \bbA') 
    = \bbI - \bbA' + \bbG,
\end{equation}
where $\bbG := \sum_{\ell=1}^{\infty} \bbA^{\ell} (\bbI - \bbA')$ is strictly lower triangular.

Substituting \eqref{eq:S} and \eqref{eq:S_inv} into \eqref{eq:equality}, we obtain
\begin{align}
    \bbD_k &= (\bbI - \bbA + \bbJ)\bbD_k'(\bbI - \bbA' + \bbG) \nonumber \\
    &= \bbD_k' + \bbD_k'(\bbG - \bbA') + (\bbJ - \bbA)\bbD_k'(\bbI - \bbA' + \bbG). \nonumber
\end{align}
The last two terms involve products of diagonal and strictly lower-triangular matrices, which remain strictly lower-triangular.
Therefore, the diagonal values of $\bbD_k$ are completely determined by those of $\bbD_k'$.
{
Recall that $[\bbD_k]_{ii}=1$ if and only if node {$v_i$} is a predecessor of node {$v_k$}, and 0 otherwise. Hence, the set $\{\bbD_k\}_{k\in\ccalV}$ uniquely encodes the ancestor sets of all nodes.
Moreover, the support of $\bbW=(\bbI-\bbA)^{-1}$ coincides with the reachability relation of the DAG, i.e., $W_{ij}\neq 0$ if and only if node {$v_j$} reaches node {$v_i$}. Therefore, $\bbW \neq \bbW'$ implies that the reachability relations of $\ccalD$ and $\ccalD'$ differ, and consequently there exists at least one node whose ancestor set is different.
This contradicts the assumption that $\bbS_k=\bbS_k'$ when $\bbW\neq\bbW'$, and the result follows.}
\hfill $\square$

%
%

\section{Proof of Theorem~\ref{thm:gf_equivariant}}\label{a:equivariance}
We first show how permutations affect the GSO $\bbS_k$ and then establish that causal graph filters are permutation equivariant.  

Permuting a graph results in a new adjacency matrix given by $\bbA' = \bbP\bbA\bbP^\top$.  
From our choice of transitive closure, it follows that the permutation reflects on $\bbW$ as  
\begin{equation}\label{eq:permuted_W}
    \bbW' = (\bbI - \bbA')^{-1} = \bbP(\bbI - \bbA)^{-1}\bbP^\top = \bbP \bbW \bbP^\top. 
\end{equation}

{Recall that the diagonal matrices $\bbD_k$ are node-associated selectors, and under a relabeling they are permuted accordingly, that is}
\begin{equation}\label{eq:permuted_D}
    \bbD_q' = \bbP\bbD_k\bbP^\top,    
\end{equation}
where {$v_k=\pi^{-1}(v_q)$}.   This ensures that $\bbD_k$ retains the ordering of the original (unpermuted) graph, while $\bbP$ reorders its diagonal entries to match the new node labels.  
Then, combining \eqref{eq:compact_gso}, \eqref{eq:permuted_W}, and \eqref{eq:permuted_D}, we obtain  
\begin{equation}
    \bbS_q' = \bbW'\bbD_q'(\bbW')^{-1} = \bbP \bbW \bbD_k \bbW^{-1} \bbP^\top = \bbP \bbS_k \bbP^\top. \nonumber
\end{equation}

Next, observe that each filter coefficient $\theta_k$ is associated with the GSO $\bbS_k$, so both are tied to a specific node rather than its label.
Thus, for each permuted term $\theta_q\bbS_q'$, we have
\begin{equation}
    \theta_q\bbS_q' = \theta_k\bbP\bbS_k\bbP^\top. \nonumber
\end{equation}
Applying this relation and the expression for causal graph filters, we conclude that the output of the graph filter when the input is given by the permuted $\bbx'$, $\bbtheta'$ and $\ccalS'$ is given by
\begin{equation}
    g(\bbx'; \bbtheta', \ccalS') = \sum_{q \in \ccalV'} \theta_q \bbS_q' \bbx' = \bbP \sum_{k \in \ccalV} \theta_k \bbS_k \bbx = \bbP g(\bbx; \bbtheta, \ccalS). \nonumber
\end{equation}
{This establishes the permutation equivariance of causal graph filters, completing the proof.}
\hfill $\square$

%
%

\section{Implementation Details}\label{app:imp_details}

We provide additional details on the implementation of the experiments reported in Section \ref{sec:experiments}; see also the GitHub repository:  \href{https://github.com/reysam93/dag_conv_nn/}{\small\texttt{https://github.com/reysam93/{dag\_conv\_nn}/}}.
In terms of computational infrastructure, experiments were conducted using Google Colab GPUs except for the more intensive experiments reported in Table \ref{t:table1}, where we utilized Tesla T4 GPUs. We first review the implementation of the proposed (P)DCN models and then present the baseline methods used for comparison. We also discuss aspects of the training procedure and hyperparameter selection. We close with the random graph models used to simulate DAGs.

\subsection*{(P)DCN Details}



{
To handle the larger number of nodes in Test Case $6$, DCN uses a two-layer architecture with a hidden dimension of $64$, while PDCN and I-PDCN use hidden dimensions of $501$ and $128$, respectively. For all other experiments reported, we employ a two-layer DCN configuration with $32$ hidden dimensions, and for PDCN a shared MLP with a single hidden layer of dimensionality $128$. These hyperparameters were selected via grid search over the hyperparameter space.

DCN-T denotes the model in which the direction of all edges in the DAG is reversed, i.e., the transposed causal GSOs are used. Finally, DCN variants that utilize only a uniformly random subset of causal GSOs, indexed by $\ccalU \subseteq \ccalV$, in the filters are denoted as DCN-$|\ccalU|$.
}



\subsection*{Baselines Methods}

We outline the baseline models and their hyperparameter used for the experimental comparisons in Section \ref{sec:experiments}.\vspace{2pt}



    
    
    

    



\noindent \textbf{GCN}~\cite{kipf2016semi}. Implements a first-order graph convolution using a degree-normalized adjacency matrix with self-loops. Each layer aggregates neighbor features and then applies a linear transformation followed by a nonlinearity. GCNs may suffer from oversmoothing. Like DCN, our implementation uses $L=2$ layers, each with $32$ learnable parameters.\vspace{2pt}

\noindent \textbf{FB-GCNN}~\cite{ruiz2021graph}. Represents the convolutional GNN architecture in \eqref{eq:fb_gnn}.
In all settings, the vector of learnable parameters has dimension $32$ and the depth is set to $L=2$ layers. Unless otherwise stated, we use a filter order $R=2$. When we vary $R$, the respective models are denoted FB-GCNN-$R$\vspace{2pt}

\noindent \textbf{GraphSAGE}~\cite{hamilton2018inductiverepresentationlearninglarge}. Learns inductive node embeddings by sampling and aggregating fixed-size neighborhoods using functions like mean, LSTM, or pooling. Unlike transductive models like GCN, it can generalize to unseen nodes after training. We implement GraphSAGE with $L=2$ layers, mean aggregation, and a hidden dimension of $32$.\vspace{2pt}

\noindent \textbf{GIN}~\cite{xu2019powerfulgraphneuralnetworks}. Designed to match the expressive power of the WL graph isomorphism test by using injective sum aggregation, thus better differentiating between graph structures. Each GIN layer combines sum-aggregated neighbor features with learnable MLPs. GIN-$L$ employs MLPs with $L$ layers. If $L$ is not specified, it defaults to $L=2$ layers. All variants use $32$ hidden units per layer. \vspace{2pt}

\noindent \textbf{GAT}~\cite{petar2018graphattentionnetworks}. Implements a self-attention mechanism to adaptively weigh the importance of neighboring nodes during feature aggregation, which is useful in graphs with heterogeneous connectivity. Each GAT layer applies multiple attention heads for a richer representation. In our setup, GAT consists of $L=2$ layers with $16$ hidden units and $2$ attention heads.\vspace{2pt}

\noindent \textbf{DAGNN}~\cite{thost2021directed}. Adapts GNN's message passing to capture a DAG's topological ordering. Uses an attention mechanism to weigh information from parent nodes and a GRU to sequentially integrate these messages. 
In our implementation, DAGNN has $L=2$ layers with hidden dimension of $128$ and dropout rate of $0.2$.\vspace{2pt}

\noindent \textbf{D-VAE}~\cite{zhang2019dvaevariationalautoencoderdirected}. A VAE model to learn continuous latent representations of DAGs. Like DAGNN, it recursively encodes nodes by aggregating messages from their predecessors using a GRU. The latent space representation is decoded back to a DAG, ensuring structural consistency. We only use D-VAE’s encoder in our tests, with a hidden layer of $128$ units.\vspace{2pt}

\noindent \textbf{MLP}~\cite{rumelhart1986learning}. A structure-agnostic model that treats node features independently, without incorporating any graph connectivity. It consists of fully connected layers applied to each node’s feature vector. Facilitates assessing the added value of using the graph structure in the models. Our MLP has $L=2$ hidden layers with $32$ neurons each.

{
\noindent \textbf{Scarselli-GNN}~\cite{scarselli2008graph}. An early GNN model that evaluates node representations through iterative fixed-point computation. Each node updates its state by aggregating messages from its neighbors until convergence. Gradients are computed via the Almeida-Pineda approach. We use a state dimension of $8$ and a maximum of $30$ iterations.\vspace{2pt}

\noindent \textbf{GRU}~\cite{cho2014learning}. A graph-agnostic recurrent baseline that treats the $N$ node signals as a sequence and processes them with a Gated Recurrent Unit. The nodes are fed in the order of their indices, which follows the topological ordering of the DAG. We use a hidden dimension of $32$.\vspace{2pt}

\noindent \textbf{LSTM}~\cite{hochreiter1997long}. Similar to GRU, this baseline treats the $N$ node signals as a sequence and processes them with a Long Short-Term Memory network. The nodes are fed in the order of their indices, which follows the topological ordering of the DAG. We use a hidden dimension of $32$.\vspace{2pt}

\noindent \textbf{DAG+NodeFormer}~\cite{luo2023transformers}. From the DAG transformer framework, this architecture augments a kernelized linear attention mechanism with DAG-aware positional encodings and an edge regularization loss that encourages attention weights to align with the true DAG edges. The attention uses random feature approximation for $O(N)$ complexity. We use $L=2$ layers with hidden dimension $32$ and $4$ attention heads.\vspace{2pt}

\noindent \textbf{DAG+SAT}~\cite{luo2023transformers}. Also from the DAG transformer framework, this architecture uses a full quadratic self-attention and adds a GCN-based structure extractor that operates on the DAG edges before the attention step. This injects local neighborhood information directly into the forward computation. We use $L=2$ layers with hidden dimension $32$ and $4$ attention heads.\vspace{2pt}}

\subsection*{Training Hyperparameters}

We specify the hyperparameters chosen to train the models used to obtain the results reported in Section \ref{sec:experiments}.\vspace{2pt}

\noindent \textbf{Learning rate.} We explored a range of learning rates, mainly between $5 \times 10^{-4}$ and $5 \times 10^{-3}$. For the River Thames and diffusion learning experiments, a learning rate of $5 \times 10^{-4}$ yielded the best results. The source identification task was trained with a learning rate of $5 \times 10^{-3}$, while gene expression imputation was performed using $10^{-3}$.\vspace{2pt}

\noindent \textbf{Batch size.} The smaller gene expression dataset dictated a batch size of $3$. Otherwise, a batch size of $25$ was used. \vspace{2pt}
    
\noindent\textbf{Number of epochs.} For gene expression imputation we trained for $500$ epochs. In all other cases we chose $100$ epochs. \vspace{2pt}
    
\noindent \textbf{Weight decay.} In all cases we used a weight decay of  $10^{-4}$.\vspace{2pt}
    
\noindent\textbf{Optimizer.} Adam~\cite{DBLP:journals/corr/KingmaB14} was used for all experiments.
    
\subsection*{Random Graph Models}

For the synthetic data experiments in Section \ref{ssec:synthetic_data}, we generate random DAGs according to the following two models.\vspace{2pt} 

\noindent \textbf{Erd\H{o}s-R\'enyi (ER).} For given $N$, undirected edges are established independently with probability $p$. Then, the edges are randomly oriented so that $\bbA$ is lower triangular.\vspace{2pt}

\noindent \textbf{Scale-Free (SF).} 
These graph models are based on the preferential attachment process~\cite{barabasi1999emergence}.  Starting from an initial DAG with \( m_0 \) nodes, new nodes are added sequentially, each connecting to \( m \leq m_0 \) existing nodes with probability proportional to their degrees. Edges are oriented each time a new node is attached, resulting in a sampled DAG. 


\bibliographystyle{IEEEtran}
\bibliography{myIEEEabrv,biblio}

\begin{thebibliography}{10}
\providecommand{\url}[1]{#1}
\csname url@samestyle\endcsname
\providecommand{\newblock}{\relax}
\providecommand{\bibinfo}[2]{#2}
\providecommand{\BIBentrySTDinterwordspacing}{\spaceskip=0pt\relax}
\providecommand{\BIBentryALTinterwordstretchfactor}{4}
\providecommand{\BIBentryALTinterwordspacing}{\spaceskip=\fontdimen2\font plus
\BIBentryALTinterwordstretchfactor\fontdimen3\font minus \fontdimen4\font\relax}
\providecommand{\BIBforeignlanguage}[2]{{%
\expandafter\ifx\csname l@#1\endcsname\relax
\typeout{** WARNING: IEEEtran.bst: No hyphenation pattern has been}%
\typeout{** loaded for the language `#1'. Using the pattern for}%
\typeout{** the default language instead.}%
\else
\language=\csname l@#1\endcsname
\fi
#2}}
\providecommand{\BIBdecl}{\relax}
\BIBdecl

\bibitem{dcn2024asilomar}
S.~Rey, H.~Ajorlou, and G.~Mateos, ``Convolutional learning on directed acyclic graphs,'' in \emph{Proc. Asilomar Conf. Signals, Syst., Computers}, 2024, pp. 423--427.

\bibitem{bronstein2017geometric}
M.~M. {Bronstein}, J.~{Bruna}, Y.~{LeCun}, A.~{Szlam}, and P.~{Vandergheynst}, ``Geometric deep learning: Going beyond {E}uclidean data,'' \emph{{IEEE} Signal Process. Mag.}, vol.~34, no.~4, pp. 18--42, July 2017.

\bibitem{ortega2018pieee}
A.~Ortega, P.~Frossard, J.~Kovačević, J.~M.~F. Moura, and P.~Vandergheynst, ``Graph signal processing: Overview, challenges, and applications,'' \emph{Proc. IEEE}, vol. 106, no.~5, pp. 808--828, 2018.

\bibitem{Dong_2020}
X.~Dong, D.~Thanou, L.~Toni, M.~Bronstein, and P.~Frossard, ``Graph signal processing for machine learning: A review and new perspectives,'' \emph{IEEE Signal Process. Mag.}, vol.~37, no.~6, p. 117–127, Nov. 2020.

\bibitem{leus2023spmag}
G.~Leus, A.~G. Marques, J.~M. Moura, A.~Ortega, and D.~I. Shuman, ``Graph signal processing: History, development, impact, and outlook,'' \emph{IEEE Signal Process. Mag.}, vol.~40, no.~4, pp. 49--60, 2023.

\bibitem{ruiz2021graph}
L.~Ruiz, F.~Gama, and A.~Ribeiro, ``Graph neural networks: Architectures, stability, and transferability,'' \emph{Proc. {IEEE}}, vol. 109, no.~5, pp. 660--682, 2021.

\bibitem{wu2020comprehensive}
Z.~Wu, S.~Pan, F.~Chen, G.~Long, C.~Zhang, and P.~S. Yu, ``A comprehensive survey on graph neural networks,'' \emph{IEEE Trans. Neural Netw. Learn. Syst.}, vol.~32, no.~1, pp. 4--24, 2021.

\bibitem{rozada2025unrolling}
S.~Rozada, S.~Rey, G.~Mateos, and A.~G. Marques, ``Unrolling dynamic programming via graph filters,'' in \emph{Proc. IEEE Intl. Wrksp. Computat. Advances Multi-Sensor Adaptive Process. (CAMSAP)}, 2025, pp. 96--100.

\bibitem{kipf2016semi}
T.~N. Kipf and M.~Welling, ``Semi-supervised classification with graph convolutional networks,'' in \emph{Proc. Int. Conf. Learn. Representations}, 2017, pp. 1--14.

\bibitem{xu2019powerfulgraphneuralnetworks}
K.~Xu, W.~Hu, J.~Leskovec, and S.~Jegelka, ``How powerful are graph neural networks?'' in \emph{Proc. Int. Conf. Learn. Representations}, 2019, pp. 1--17.

\bibitem{petar2018graphattentionnetworks}
P.~Veli{\v{c}}kovi{\'{c}}, G.~Cucurull, A.~Casanova, A.~Romero, P.~Li{\`{o}}, and Y.~Bengio, ``Graph attention networks,'' in \emph{Proc. Int. Conf. Learn. Representations}, 2018, pp. 1--12.

\bibitem{wang2017mgae}
C.~Wang, S.~Pan, G.~Long, X.~Zhu, and J.~Jiang, ``{MGAE}: Marginalized graph autoencoder for graph clustering,'' in \emph{Assoc. Comput. Mach.}, 2017, pp. 889--898.

\bibitem{rey2021overparametrized}
S.~Rey, V.~M. Tenorio, S.~Rozada, L.~Martino, and A.~G.~Marques, ``Overparametrized deep encoder-decoder schemes for inputs and outputs defined over graphs,'' in \emph{Proc. European Signal Process. Conf. (EUSIPCO)}.\hskip 1em plus 0.5em minus 0.4em\relax IEEE, 2021, pp. 855--859.

\bibitem{scarselli2008graph}
F.~Scarselli, M.~Gori, A.~C. Tsoi, M.~Hagenbuchner, and G.~Monfardini, ``The graph neural network model,'' \emph{{IEEE} Trans. Neural Netw. Learning Syst.}, vol.~20, no.~1, pp. 61--80, 2008.

\bibitem{marques2020signal}
A.~G. Marques, S.~Segarra, and G.~Mateos, ``Signal processing on directed graphs: The role of edge directionality when processing and learning from network data,'' \emph{{IEEE} Signal Process. Mag.}, vol.~37, no.~6, pp. 99--116, 2020.

\bibitem{seifert2023causal}
B.~Seifert, C.~Wendler, and M.~Püschel, ``Causal {F}ourier analysis on directed acyclic graphs and posets,'' \emph{IEEE Trans. Signal Process.}, vol.~71, pp. 3805--3820, 2023.

\bibitem{huang2022graph}
C.~Huang, M.~Li, F.~Cao, H.~Fujita, Z.~Li, and X.~Wu, ``Are graph convolutional networks with random weights feasible?'' \emph{{IEEE} Trans. Pattern Anal. Mach. Intell.}, vol.~45, no.~3, pp. 2751--2768, 2022.

\bibitem{yang2025deeper}
G.~Yang, M.~Li, H.~Feng, and X.~Zhuang, ``Deeper insights into deep graph convolutional networks: Stability and generalization,'' \emph{{IEEE} Trans. Pattern Anal. Mach. Intell.}, 2025.

\bibitem{peters2017elements}
J.~Peters, D.~Janzing, and B.~Sch{\"o}lkopf, \emph{Elements of Causal Inference: Foundations and Learning Algorithms}.\hskip 1em plus 0.5em minus 0.4em\relax The MIT Press, 2017.

\bibitem{peters2014identifiability}
J.~Peters and P.~B{\"u}hlmann, ``Identifiability of {G}aussian structural equation models with equal error variances,'' \emph{Biometrika}, vol. 101, no.~1, pp. 219--228, 2014.

\bibitem{zheng2018dags}
X.~Zheng, B.~Aragam, P.~K. Ravikumar, and E.~P. Xing, ``{DAG}s with no tears: Continuous optimization for structure learning,'' \emph{Proc. Adv. Neural. Inf. Process. Syst.}, vol.~31, 2018.

\bibitem{saboksayr2023colide}
S.~S. Saboksayr, G.~Mateos, and M.~Tepper, ``{CoLiDE}: Concomitant linear {DAG} estimation,'' in \emph{Proc. Int. Conf. Learn. Representations}, 2024.

\bibitem{dAcunto2023multiscale}
G.~D'Acunto, P.~D. Lorenzo, and S.~Barbarossa, ``Multiscale causal structure learning,'' \emph{Trans. Mach. Learn. Res.}, pp. 1--39, 2023.

\bibitem{rey2025non}
S.~Rey, S.~S. Saboksayr, and G.~Mateos, ``Non-negative weighted dag structure learning,'' in \emph{Proc. IEEE Intl. Conf. Acoustics, Speech and Signal Process. (ICASSP)}.\hskip 1em plus 0.5em minus 0.4em\relax IEEE, 2025, pp. 1--5.

\bibitem{allamanis2018survey}
M.~Allamanis, E.~T. Barr, P.~Devanbu, and C.~Sutton, ``A survey of machine learning for big code and naturalness,'' \emph{ACM Computing Surveys (CSUR)}, vol.~51, no.~4, pp. 1--37, 2018.

\bibitem{zhang2018graph}
C.~Zhang, M.~Ren, and R.~Urtasun, ``Graph hypernetworks for neural architecture search,'' in \emph{Proc. Int. Conf. Learn. Representations}, 2019.

\bibitem{puschel2021discrete}
M.~Püschel, B.~Seifert, and C.~Wendler, ``Discrete signal processing on meet/join lattices,'' \emph{{IEEE} Trans. Signal Process.}, vol.~69, pp. 3571--3584, 2021.

\bibitem{zhang2019dvaevariationalautoencoderdirected}
M.~Zhang, S.~Jiang, Z.~Cui, R.~Garnett, and Y.~Chen, ``{D-VAE}: A variational autoencoder for directed acyclic graphs,'' in \emph{Proc. Adv. Neural. Inf. Process. Syst.}, 2019.

\bibitem{thost2021directed}
V.~Thost and J.~Chen, ``Directed acyclic graph neural networks,'' in \emph{Int. Conf. Learn. Representations}, 2021.

\bibitem{WL1968test}
B.~Y. Weisfeiler and A.~A. Lehman, ``A reduction of a graph to a canonical form and an algebra arising during this reduction,'' \emph{Nauchno-Technicheskaya Informatsia}, vol.~2, no.~9, pp. 12--16, 1968.

\bibitem{luo2023transformers}
Y.~Luo, V.~Thost, and L.~Shi, ``Transformers over directed acyclic graphs,'' in \emph{Proc. Adv. Neural. Inf. Process. Syst.}, vol.~36, 2023, pp. 47\,764--47\,782.

\bibitem{elvin2024gf}
E.~Isufi, F.~Gama, D.~I. Shuman, and S.~Segarra, ``Graph filters for signal processing and machine learning on graphs,'' \emph{IEEE Trans. Signal Process.}, vol.~72, pp. 4745--4781, 2024.

\bibitem{rey2025redesigning}
S.~Rey, M.~Navarro, V.~M. Tenorio, S.~Segarra, and A.~G. Marques, ``Redesigning graph filter-based {GNN}s to relax the homophily assumption,'' in \emph{Proc. IEEE Intl. Conf. Acoustics, Speech and Signal Process. (ICASSP)}, 2025, pp. 1--5.

\bibitem{lehmann1977algebraic}
D.~J. Lehmann, ``Algebraic structures for transitive closure,'' \emph{Theoretical Comput. Sci.}, vol.~4, no.~1, pp. 59--76, 1977.

\bibitem{rota1964foundations}
G.-C. Rota, ``On the foundations of combinatorial theory i. theory of m{\"o}bius functions,'' \emph{Probability Theory and Related Fields}, vol.~2, pp. 340--368, 1964.

\bibitem{rey2022untrained}
S.~Rey, S.~Segarra, R.~Heckel, and A.~G. Marques, ``Untrained graph neural networks for denoising,'' \emph{{IEEE} Trans. Signal Process.}, vol.~70, pp. 5708--5723, 2022.

\bibitem{peng2024beyond}
J.~Peng, R.~Lei, and Z.~Wei, ``Beyond over-smoothing: Uncovering the trainability challenges in deep graph neural networks,'' in \emph{Proc. 33rd ACM Int. Conf. Inf. Knowl. Manag.}, 2024, pp. 1878--1887.

\bibitem{arroyo2025vanishing}
A.~Arroyo, A.~Gravina, B.~Gutteridge, F.~Barbero, C.~Gallicchio, X.~Dong, M.~Bronstein, and P.~Vandergheynst, ``On vanishing gradients, over-smoothing, and over-squashing in gnns: Bridging recurrent and graph learning,'' \emph{arXiv preprint arXiv:2502.10818}, 2025.

\bibitem{doshi2022graph}
S.~Doshi and S.~P. Chepuri, ``Graph neural networks with parallel neighborhood aggregations for graph classification,'' \emph{IEEE Trans. Signal Process.}, vol.~70, pp. 4883--4896, 2022.

\bibitem{hamilton2018inductiverepresentationlearninglarge}
W.~L. Hamilton, R.~Ying, and J.~Leskovec, ``Inductive representation learning on large graphs,'' in \emph{Proc. Adv. Neural. Inf. Process. Syst.}, 2017, pp. 1025--1035.

\bibitem{li2016gated}
Y.~Li, R.~Zemel, M.~Brockschmidt, and D.~Tarlow, ``Gated graph sequence neural networks,'' in \emph{Proc. Int. Conf. Learn. Representations}, 2016.

\bibitem{rumelhart1986learning}
D.~E. Rumelhart, G.~E. Hinton, and R.~J. Williams, ``Learning representations by back-propagating errors,'' \emph{Nature}, vol. 323, no. 6088, pp. 533--536, 1986.

\bibitem{lstm}
S.~Hochreiter and J.~Schmidhuber, ``Long short-term memory,'' \emph{Neural Comput.}, vol.~9, pp. 1735--1780, 11 1997.

\bibitem{gru}
K.~Cho, B.~van Merri{\"e}nboer, D.~Bahdanau, and Y.~Bengio, ``On the properties of neural machine translation: Encoder{--}decoder approaches,'' in \emph{Proc. Wkshp. Syntax, Semantics and Structure in Statistical Translation}, Doha, Qatar, Oct. 2014, pp. 103--111.

\bibitem{barabasi1999emergence}
A.-L. Barab{\'a}si and R.~Albert, ``Emergence of scaling in random networks,'' \emph{Science}, vol. 286, no. 5439, pp. 509--512, 1999.

\bibitem{opgenrhein2007correlation}
R.~Opgen-Rhein and K.~Strimmer, ``From correlation to causation networks: A simple approximate learning algorithm and its application to high-dimensional plant gene expression data,'' \emph{BMC Systems Biology}, vol.~1, no.~1, p.~37, 2007.

\bibitem{bowes2020weekly}
M.~J. Bowes, L.~K. Armstrong, S.~A. Harman, D.~J.~E. Nicholls, H.~D. Wickham, P.~M. Scarlett, and M.~D. Juergens, ``Weekly water quality data from the {R}iver {T}hames and its major tributaries (2009–2017),'' 2020.

\bibitem{cho2014learning}
K.~Cho, B.~van Merrienboer, C.~Gulcehre, D.~Bahdanau, F.~Bougares, H.~Schwenk, and Y.~Bengio, ``Learning phrase representations using {RNN} encoder-decoder for statistical machine translation,'' in \emph{Proc. 2014 Conf. Empirical Methods in Natural Language Processing (EMNLP)}, 2014.

\bibitem{hochreiter1997long}
S.~Hochreiter and J.~Schmidhuber, ``Long short-term memory,'' \emph{Neural computation}, vol.~9, no.~8, pp. 1735--1780, 1997.

\bibitem{DBLP:journals/corr/KingmaB14}
D.~P. Kingma and J.~Ba, ``Adam: {A} method for stochastic optimization,'' in \emph{Proc. Int. Conf. Learn. Representations}, 2015.

\end{thebibliography}


\begin{IEEEbiography}[{\includegraphics[width=1in,height=1.80in,clip,keepaspectratio]{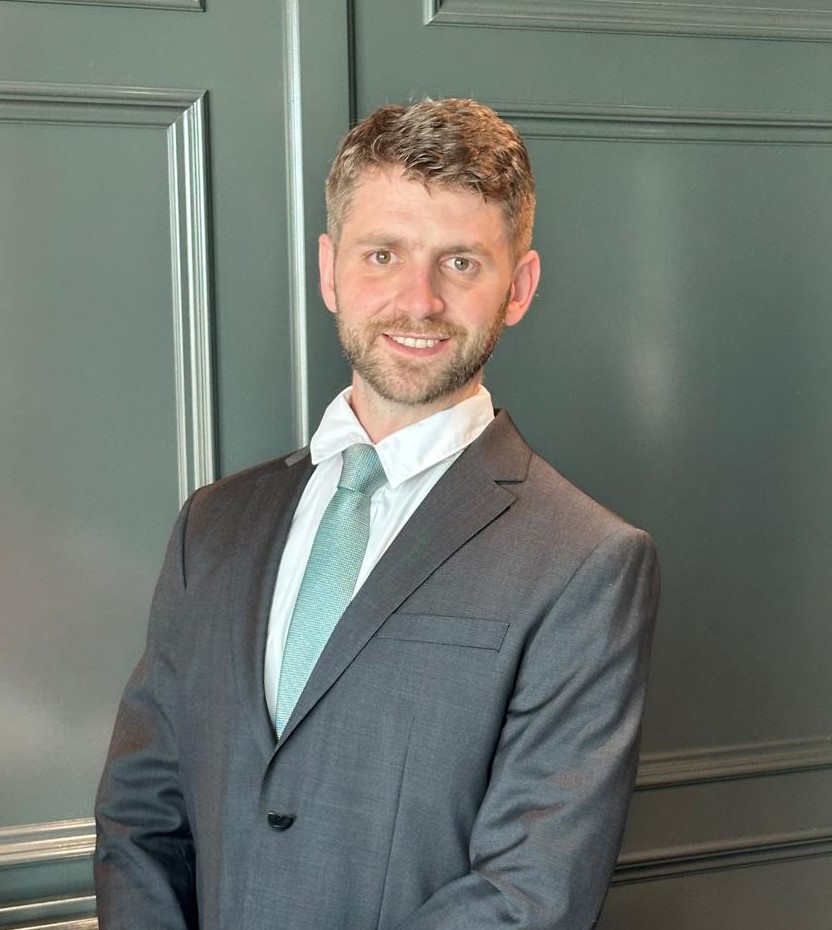}}]
{Samuel Rey (Member, IEEE)} received the B.Sc., M.Sc., and Ph.D. degrees in telecommunication engineering from King Juan Carlos University (URJC), Madrid, Spain, in 2016, 2018, and 2023, respectively, all with highest honors. In 2023, he joined the Department of Signal Theory and Communications, URJC, where he is currently an Assistant Professor. 

His current research interests include graph signal processing, machine learning, nonconvex optimization, and data science over networks. Dr. Rey has served on the organizing committees of several international conferences. He received the Best Young Investigator Award from URJC in 2018 and was awarded the Spanish National FPU Scholarship for PhD studies in the same year.
\end{IEEEbiography}

\begin{IEEEbiography}[{\includegraphics[width=1in,height=1.80in,clip,keepaspectratio]{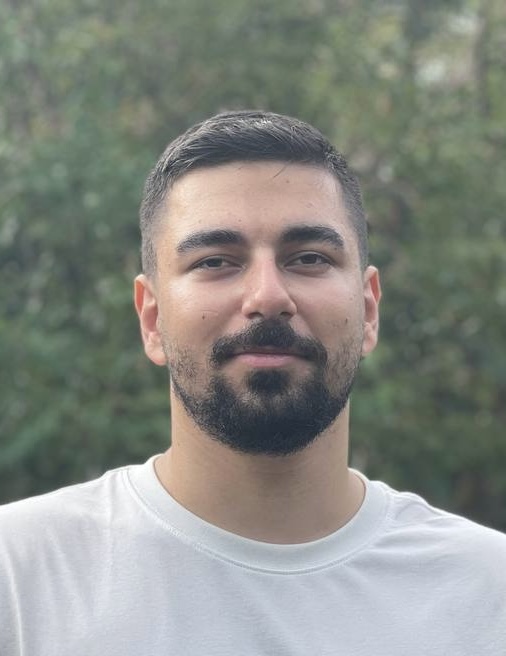}}]{Hamed Ajorlou (Student Member, IEEE)} received the B.Sc. degree in electrical engineering from Sharif University of Technology, Tehran, Iran, in 2023 and the M.Sc. degree in electrical engineering from the University of Rochester, Rochester, NY, USA, in 2025. Since 2023, he has been working toward the Ph.D. degree at the University of Rochester. His research interests are in developing efficient algorithms for large-scale graph processing and exploring the theoretical foundations of graph neural networks.
\end{IEEEbiography}

\begin{IEEEbiography}[{\includegraphics[width=1in,height=1.80in,clip,keepaspectratio]{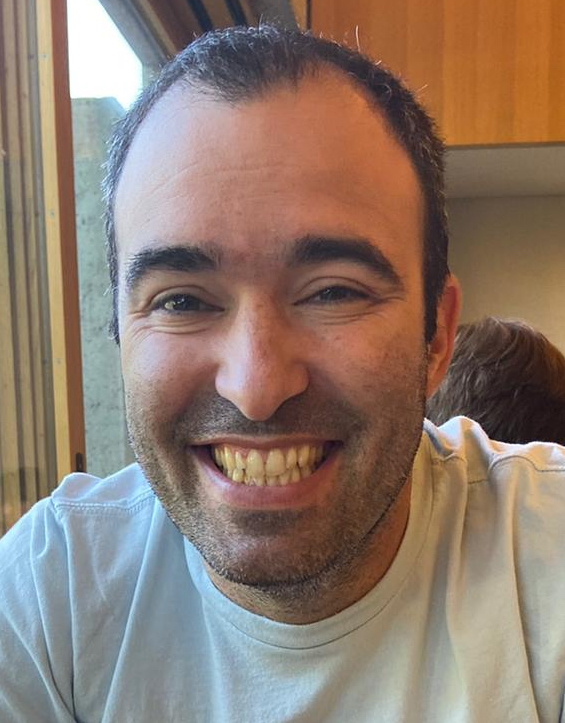}}]{Gonzalo Mateos (Senior Member, IEEE)} received his B.Sc. degree in electrical engineering from Universidad de la Rep\'ublica, Montevideo, Uruguay in 2005, and the M.Sc. and Ph.D. degrees in electrical
engineering from the University of Minnesota, Minneapolis, MN, USA, in 2009 and 2012, respectively. From 2004 to 2006, he was a Systems Engineer
with Asea Brown Boveri, Uruguay. In 2013, he was
a Visiting Scholar with the Computer Science Department, Carnegie Mellon University, Pittsburgh, PA, USA. In 2014, he joined the University of Rochester, Rochester, NY, USA, where he is currently a Professor with the Department of Electrical and Computer Engineering, the Department of Computer Science (secondary appointment), and the Associate Director for Research with the Goergen Institute for Data Science and Artificial Intelligence, University of Rochester. His research interests include statistical learning from complex data, network science, decentralized optimization, causal discovery, and graph signal processing. 	
\end{IEEEbiography}

\end{document}